\documentclass[times,twocolumn]{aastex62}

\usepackage{cases}
\usepackage{graphicx}
\usepackage{CJK}
\usepackage{subfigure}
\usepackage{natbib}
\usepackage{color}
\usepackage{tabularx}
\usepackage{bm}
\usepackage{threeparttable}
\usepackage{lmodern}
\usepackage{aas_macros}
\usepackage{gensymb}
\usepackage[T1]{fontenc}
\usepackage[utf8]{inputenc}

\usepackage{amsmath}
\usepackage{booktabs}

\newcommand{\aemb}{{\sc Aemulus}-$\beta$}
\newcommand{\um}{{\sc UniverseMachine}}

\begin{document}

\preprint{APS/123-QED}

\title{Stress-Testing Small-Scale Cosmological Inference With Empirical Galaxy--Halo Connection Models}

\author{Hanyue Wang}
\affiliation{Department of Physics, Stanford University, 382 Via Pueblo Mall, Stanford, CA 94305, USA}
\affiliation{Kavli Institute for Particle Astrophysics and Cosmology, 452 Lomita Mall, Stanford, CA 94305, USA}
\affiliation{SLAC National Accelerator Laboratory, Menlo Park, CA 94025, USA}
\author{Kate Storey-Fisher}
\affiliation{Kavli Institute for Particle Astrophysics and Cosmology, 452 Lomita Mall, Stanford, CA 94305, USA}
\affiliation{SLAC National Accelerator Laboratory, Menlo Park, CA 94025, USA}
\author{Philip Mansfield}
\affiliation{Kavli Institute for Particle Astrophysics and Cosmology, 452 Lomita Mall, Stanford, CA 94305, USA}
\author{Andrew Hearin}
\affiliation{HEP Division, Argonne National Laboratory, 9700 South Cass Avenue, Lemont, IL 60439, USA}
\author{Marcelo Alvarez}
\affiliation{Kavli Institute for Particle Astrophysics and Cosmology, 452 Lomita Mall, Stanford, CA 94305, USA}
\affiliation{SLAC National Accelerator Laboratory, Menlo Park, CA 94025, USA}
\author{Suchetha Cooray}
\affiliation{Astronomical Institute, Tohoku University, 6-3, Aramaki, Aoba-ku, Sendai, Miyagi, 980-8578, Japan}
\author{Risa H. Wechsler}
\affiliation{Department of Physics, Stanford University, 382 Via Pueblo Mall, Stanford, CA 94305, USA}
\affiliation{Kavli Institute for Particle Astrophysics and Cosmology, 452 Lomita Mall, Stanford, CA 94305, USA}
\affiliation{SLAC National Accelerator Laboratory, Menlo Park, CA 94025, USA}

\shorttitle{Cosmological inference with small-scale clustering}
\shortauthors{H. Wang et al.}

\begin{abstract}
Small-scale galaxy clustering contains substantial cosmological information, but extracting it reliably requires robust modeling of the galaxy--halo connection. In this work, we compare two widely adopted families of empirical models, subhalo abundance matching (SHAM) and the halo occupation distribution (HOD), testing their predictions for galaxy clustering and their recovery of the input cosmology.
Using \aemb, a new multi-cosmology simulation suite with substructure merger trees, we fit multiple SHAM and HOD models to the small-scale clustering of a stellar-mass-selected mock galaxy catalog generated with the more physical semi-empirical model \um. 
We then use cosmological evidence modeling to infer the structure growth parameter $S_8$. 
Our results show that the SHAM model, regardless of the treatment of numerically disrupted subhalos, is able to offer tight, unbiased constraints on $S_8$ with $\sim 2$--$3\%$ precision, using simulations of relatively small volumes that span a limited number of cosmologies.
In contrast, HOD models, even when accounting for assembly bias, yield weaker and potentially biased cosmological constraints that are sensitive to implementation choices: the inferred $S_8$ degrades in precision to $\sim 6$--$11\%$ and can be offset by $\sim 1\sigma$.
This suggests that the excessive parametric flexibility and model misspecification of most commonly adopted HOD models can degrade and bias the resulting cosmological constraints.
We further evaluate each model's performance in predicting higher-order statistics not used in the fit and find that SHAM generally reproduces them more accurately than the tested HOD models.
These results highlight the potential of leveraging predictive subhalo-level information from simulations to enable robust cosmological inference on small scales.
\end{abstract}


\medskip

\section{Introduction}

Understanding the full information content in the distribution of galaxies is one of the main goals of modern astrophysics and cosmology. With the new generation of large sky surveys, such as the Dark Energy Spectroscopic Instrument \citep[DESI,][]{DESI1, DESI2}, the Euclid mission \citep{Euclid}, the Roman Space Telescope \citep{Roman}, and the Rubin Observatory Legacy Survey of Space and Time \citep[LSST,][]{LSST}, we are entering an era of high-precision studies of galaxy clustering across a wide range of scales with galaxy distribution maps of unprecedented volume and density.

To fully exploit these datasets and enhance cosmological constraints, extracting information from small, non-linear scales is essential. These scales contain a large number of measurable modes and can therefore add substantial constraining power \citep[e.g.,][]{eifler_krause_2017_cosmolike,zhai_etal19_aemulus3}. However, since analytic perturbation theory breaks down in this regime, dark matter simulations together with robust modeling of the relationship between galaxies and halos \citep{Wechsler&Tinker} are necessary to make accurate and robust predictions for relevant statistics. Such models need to be predictive enough to include the full cosmological information, accurate enough to characterize the dependence of the spatial distribution on key observables, and flexible enough to marginalize over uncertainties in galaxy formation.

While hydrodynamical simulations (e.g. \citealt{Illustris, EAGLE, Flamingo}) directly model the physical processes of galaxy formation, including baryonic effects, they remain computationally too expensive for iterative evaluations across cosmologies. Semi-analytic models (e.g. \citealt{Somerville2008, Somerville2015, Henrique2015, Henrique2020, Lagos2024}) and semi-empirical forward models (e.g. \citealt{EMERGE, UniverseMachine, Diffstar}) aim to approximate these physical processes more efficiently and can be calibrated against observational data. However, the large number of model parameters has so far made it difficult to explore a broad galaxy formation parameter space, and the computational cost of constructing these models over large cosmological volumes remains nontrivial for inference applications. 
Simple empirical galaxy--halo connection models are therefore widely used for their low computational cost and their flexibility to fit observational data. 


Two of the most widely adopted empirical frameworks for the galaxy--halo connection are the subhalo abundance matching model (SHAM) and the halo occupation distribution model (HOD). 
SHAM assigns galaxy properties by ranking (sub)halos according to a chosen halo property and matching their abundances to those of galaxies, generally allowing scatter in the relation \citep{Kravtsov2004, Vale2006, Conroy2006, Reddick2013}.
HOD models instead specify the probability distribution of galaxy occupation as a parametric function of host halo properties, usually beginning with halo mass \citep{Peacock2000, Seljak2000, BWS2002, BerlindWeinberg2002, Zheng2005}. 
Standard HOD models depend only on host-halo mass and do not explicitly model how occupation at fixed mass varies with secondary halo properties such as formation history or concentration, an effect known as assembly bias \citep{Wechsler2001, Gao2005, Wechsler2006, Mao2018}.
Such dependencies have been incorporated into HOD models more commonly in recent years (e.g. \citealt{Hearin2016, Xu2021, AbacusHOD}).

SHAM makes use of more information from the simulation, such as the properties and assembly histories of subhalos, and naturally incorporates assembly bias, while HOD models rely only on host-halo catalogs and are thus less computationally demanding. Both empirical models have been broadly employed to fit observational data and study properties of different galaxy samples, and have demonstrated success in reproducing clustering statistics. However, it remains unclear whether they recover cosmology equally reliably when small-scale measurements are included, and how effectively each can properly marginalize over uncertainties in the galaxy--halo connection.

In this work, we investigate requirements for galaxy--halo connection modeling within these empirical frameworks in order to support cosmological inference on small scales. Using \aemb, a new multi-cosmology simulation suite with sufficient resolution for reliable merger trees and subhalo catalogs, we stress-test multiple SHAM and HOD models against a realistic mock galaxy catalog constructed with the more complex semi-empirical model \um\ \citep{UniverseMachine}, which provides a physically motivated and observationally grounded galaxy--halo connection. 
At a fixed fiducial cosmology, we evaluate the accuracy of each model in reproducing two-point clustering down to small scales. We then investigate whether the modeling frameworks can be employed to infer unbiased cosmological parameters using the Cosmological Evidence Modeling \citep[CEM,][]{CEM} method, by fitting the target clustering with models built on all \aemb\ cosmologies. 
The high resolution of \aemb\ allows SHAM to be built directly on simulated substructures across different cosmologies for the first time, enabling subhalo-level information to be propagated into the cosmological inference.
As an additional predictive test, we compare each model’s predictions for two higher-order statistics not used in the fits, counts-in-cylinders and multiplet clustering, with actual measurements from the target catalog.

\begin{figure*}
    \centering
    \includegraphics[width=1.9\columnwidth]{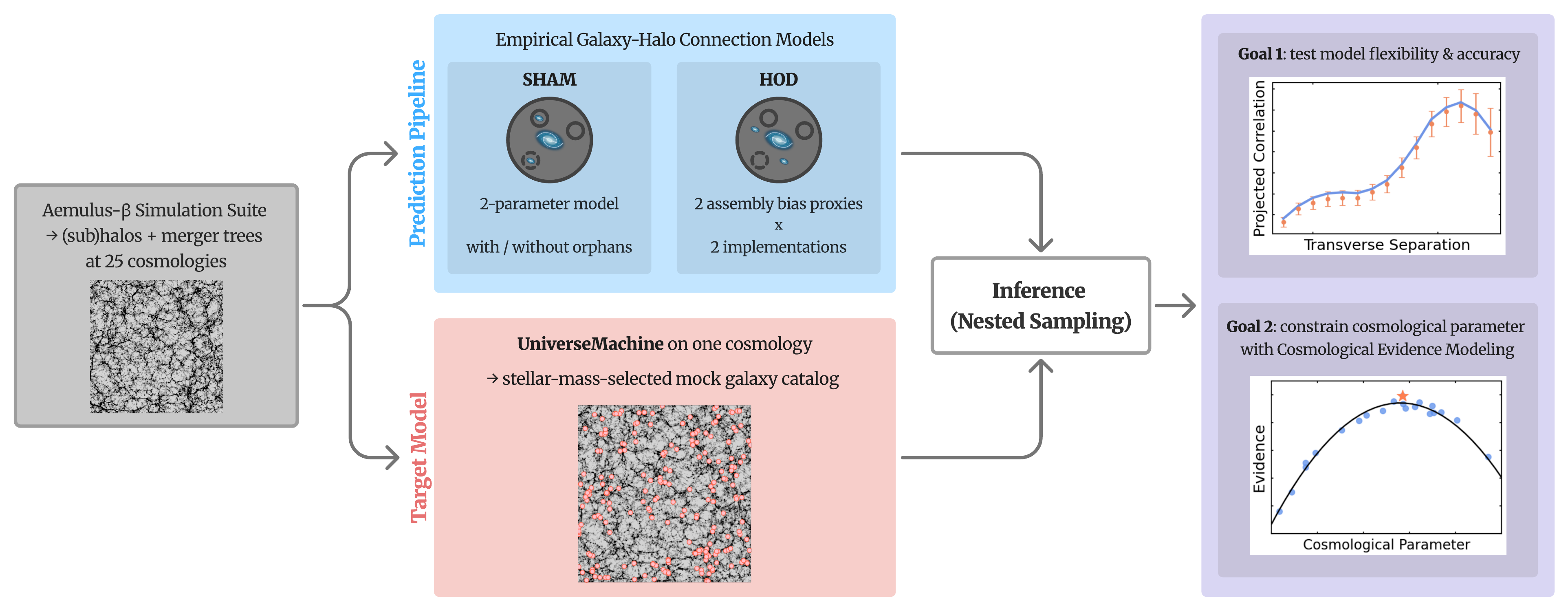}
    \caption{Schematic overview of the workflow of this paper. Starting from the \aemb\ simulation suite, we construct a target mock galaxy catalog using \um, and build two classes of empirical galaxy–halo connection models, SHAM and HOD, as prediction pipelines. We then fit the empirical models to the clustering of the target catalog via nested sampling, testing their ability to describe the underlying galaxy–halo connection and offer unbiased cosmological constraints.}
    \label{fig: outline}
\end{figure*}

This paper is structured as follows. Section~\ref{sec: sim} describes the \aemb\ simulation suite and the target mock galaxy catalog constructed using \um. Section~\ref{sec: model} introduces the SHAM and HOD models, and Section~\ref{sec: method} the statistical methods. 
In Section~\ref{sec: fit}, we examine assembly bias in the \um\ mock and present model fits at fixed cosmology. In Section~\ref{sec: cos}, we present and discuss cosmological constraints from different models. We extend the analysis to higher-order statistics in Section~\ref{sec: acm}. We discuss limitations and implications in Section~\ref{sec: dis} and conclude in Section~\ref{sec: conclusion}.
A schematic overview of the workflow of the paper is shown in Fig.~\ref{fig: outline}.

\section{Simulations and Galaxy Catalogs} \label{sec: sim}

\subsection{Aemulus-$\beta$ Simulations}

We use dark matter catalogs from the \aemb\ simulation suite (Alvarez~et~al.~2026, in prep) to construct various empirical galaxy--halo connection models. The \aemb\ suite consists of 25 simulations that span a 7-dimensional $w$CDM cosmological parameter space: the baryon density $\omega_{b}$, the cold dark matter density $\omega_{c}$, the Hubble constant $H_{0}$, the spectral index $n_{s}$, the amplitude of scalar perturbations $A_{s}$, the dark energy equation of state $w$, and the effective number of neutrino species $N_{\rm eff}$. The 25 cosmologies span the same design volume as the original {\sc Aemulus} suite \citep{2019ApJ...875...69D}: a Latin hypercube constructed in the eigenbasis of the parameter covariance of combined WMAP9 and Planck 2013 chains with BOSS DR11 BAO and Type Ia supernova measurements, with each hypercube dimension spanning $\pm4\sigma$ along its corresponding eigenvector.

Each simulation contains $2048^{3}$ particles with a particle mass of $m_p \approx 6.2 \times 10^8(\Omega_m/0.3)\,h^{-1}\,M_\odot$ in a periodic volume of $(400\,h^{-1}\rm Mpc)^{3}$. The Plummer-equivalent softening length is $3\,{\rm kpc}~h^{-1}$, corresponding to roughly 0.015 times the mean-interparticle spacing. This high resolution allows us to construct subhalo catalogs and merger trees over a broad range of different cosmologies, enabling cosmological analysis with reliable small-scale information. 
Dark matter halos are identified using the \textsc{Rockstar} halo finder \citep{Rockstar}, and merger trees are constructed using \textsc{consistent-trees} \citep{ctree}. We use the $z=0.07$ redshift snapshots throughout the work. Of the 25 \aemb\ simulations, currently 22 have usable catalogs and merger trees; the remaining are excluded due to corrupted data.

\subsection{UniverseMachine \& the Mock Galaxy Catalog}

We use \um\ \citep{UniverseMachine} to construct a physically motivated, observationally grounded mock galaxy catalog against which to test the SHAM and HOD models.
\um\ ties star formation directly to halo assembly history and is well-calibrated to observations, making it a plausible realization of the real universe with substantially more complex galaxy--halo connection than the models we test.
Specifically, \um\ assigns star formation rates (SFRs) to individual galaxies using the maximum circular velocities, assembly histories, and redshifts of their halos, and then integrates these SFRs along merger trees to self-consistently predict the galaxy stellar masses.
Thereafter, \um\ forward models a comprehensive list of summary statistics, including stellar mass functions, specific SFRs, quenched fractions, UV luminosity functions, and clustering. The predicted statistics are then compared with observations to calibrate the model parameters.

We populate one \aemb\ simulation with the \um\ Data Release 1 model to construct the target mock galaxy catalog. Building the test and target models on the same simulation suite ensures that they share the mass resolution and halo identification. 
We choose to generate the target mock at a cosmology close to Planck 2018 \citep{Planck2018} and near the center of the parameter space spanned by \aemb. The final mock galaxy catalog is a stellar-mass-selected sample with $\log(M_\star/M_\odot) \geq 10.5$, yielding a number density of $n_g \sim 2\times10^{-3}\,{\rm Mpc}^{-3}$.
This number density is comparable to a volume-limited DESI BGS-like sample to $z\sim 0.2$ (estimated by integrating the observed luminosity function). 
A denser sample would improve access to small-scale information, but we do not adopt this for two reasons: first, in real observations a lower stellar mass threshold restricts the volume over which the sample remains complete, leaving the clustering measurements cosmic-variance limited; and second, we set the threshold conservatively above the mass scale at which the \aemb\ resolution begins to affect the subhalo population. For validation tests at fixed simulation volume, we would expect cosmological constraints to broaden mildly for lower-density samples.

\section{Galaxy--Halo Connection Models} \label{sec: model}

We populate the simulated halo catalogs with two families of empirical models: subhalo abundance matching (SHAM, \S \ref{sec: SHAM}), which uses individual (sub)halo properties, and the halo occupation distribution (HOD, \S \ref{sec: HOD}), which assigns galaxies statistically to host halos.

\subsection{Subhalo Abundance Matching} \label{sec: SHAM}
The SHAM framework assumes a monotonic relationship between a galaxy proxy and a halo proxy, with some scatter in the galaxy--halo relation. Here we adopt a log-normal scatter $s$ in stellar mass at fixed halo proxy.
Building on the interpolation method introduced in \cite{Lehmann2017}, we propose an extended two-parameter SHAM model that uses a generalized halo proxy
\begin{equation}
    v_{\alpha} = v_{\rm vir}'\left(\frac{v_{\rm vir}}{v_{\rm vir}'}\right)^{\alpha},
\end{equation}
where $v_{\rm vir}$ is the virial velocity evaluated at each halo's epoch of peak mass $a_{M_{\rm peak}}$, and $v_{\rm vir}'$ is equivalent to a unit conversion of the peak halo mass $M_{\rm peak}$ using the present-day virial density threshold, as defined in Eq.~\ref{Eq: actual} and Eq.~\ref{Eq: L17}.
This parametrization combines the mass proxy $M_{\rm peak}$ with an adjustable dependence on when each halo reaches its peak mass through the additional parameter $\alpha$, as the ratio $v_{\rm vir}/v_{\rm vir}'$ depends only on $a_{M_{\rm peak}}$.
Since subhalos tend to reach $M_{\rm peak}$ earlier, this model effectively tunes the satellite fraction of the resulting galaxy sample. 
When $\alpha=0$, ranking with $v_{\alpha=0}$ is equivalent to ranking with $M_{\rm peak}$, one of the most popular halo proxies for SHAM; as $\alpha$ increases, earlier-peaking objects, typically subhalos, are ranked higher, increasing the satellite fraction and the clustering amplitude.
This halo proxy parameterization provides a smooth interpolation between ranking by $M_{\rm peak}$ and ranking by $v_{\text{vir}}(a_{M_{\rm peak}})$, spanning a range that encompasses widely adopted halo proxies such as $v_{\rm peak}$.
Appendix~\ref{app1} discusses this parametrization further and compares it with the original proxy of \cite{Lehmann2017}.

One complexity regarding SHAM models is the treatment of systems that, due to tidal stripping, fall below the resolution limit or the tracking limit of the halo finder after accretion, commonly referred to as ``orphan satellites''. 
In such systems, an associated satellite galaxy could be present even after the subhalo is lost.
The treatment of orphans has been shown to affect predictions from semi-analytic, semi-empirical, and SHAM models, with the effect depending on simulation resolution and halo finder (e.g., \citealt{Guo&White2014, UniverseMachine, SHAMe}). 
A variety of orphan prescriptions exist in the literature, but many have not been carefully tested for convergence or for consistency with the full simulation dynamics. Orphan modeling thus introduces additional modeling uncertainty.

In this work, we adopt the galaxy disruption model of \um\ to incorporate orphans when building the SHAM model described above. \um\ follows subhalos until they lose enough mass to drop out of the halo catalog. 
At this point, subhalos are replaced by massless orphans which are evolved in idealized approximations of their hosts' potentials. Along these orbits, orphans experience mass loss according to a population-averaged prescription and evolve $v_{\rm max}$ according to simulated tidal tracks. Orphans are removed from the simulation once $v_{\rm max}$ falls below a disruption threshold relative to its infall value. See Appendix B of \citet{UniverseMachine} for details. 
To investigate the impact of orphans, we treat the inclusion of orphans as a modeling choice to be tested rather than fixed: we construct the SHAM model both on the raw \textsc{Rockstar} halo catalog and on the modified halo catalog that incorporates orphans, and fit each to the target clustering statistics. We discuss the performance of both SHAM models in fitting the target clustering and inferring cosmology respectively in Section~\ref{sec: fit} and Section~\ref{sec: cos}. 

\subsection{Halo Occupation Distribution} 
\label{sec: HOD}
The HOD framework is a probabilistic model that describes occupation statistics, namely the mean number of galaxies per halo and its moments, assuming that each halo can host at most one central galaxy and some unbounded number of satellite galaxies.
Here we use the ``vanilla'' HOD model, which depends only on halo mass, following \cite{Zheng07}:
\begin{equation}
    \bar{n}_{\rm cent}(M)=\frac{1}{2}\left(1+ \text{erf}\,\left [\frac{\log_{10}(M/M_{\rm cut})}{\sqrt{2}\sigma}\right]\right),
\end{equation}
\begin{equation}
    \bar{n}_{\rm sat}(M)=\left[\frac{M-\kappa M_{\rm cut}}{M_{1}}\right]^{\alpha_{\rm HOD}} \bar{n}_{\rm cent}(M),
\end{equation}
with five free parameters $M_{\rm cut}$, $M_{1}$, $\sigma$, $\alpha_{\rm HOD}$, and $\kappa$. In this parametrization, the central galaxy occupation follows a Bernoulli distribution and the satellite occupation follows a Poisson distribution. 

Galaxy occupation can depend on secondary halo properties at fixed halo mass \citep[assembly bias;][]{Wechsler2001, Gao2005, Wechsler2006, Mao2018}, and this effect can bias inference based on mass-only HOD models \citep[e.g.,][]{zentner_etal14_assembias}.
We therefore extend the vanilla HOD to include galaxy assembly bias; and to investigate the impact of different modeling approaches, we implement this extension using two widely adopted formulations: the decorated HOD (dHOD) formalism \citep{Hearin2016} and the AbacusHOD formalism, which is developed to meet the mock requirements of DESI \citep{AbacusHOD}.

For each HOD formalism, we test two secondary properties, halo concentration and local environment, and always consider them separately.
Halo concentration is a natural choice since it encodes information about halo assembly history and is strongly correlated with halo age \citep{Wechsler2002, Wechsler2006}. 
The choice of local environment is motivated by recent studies of hydrodynamic simulations and observations, which show that considering the local dark matter overdensity can improve the fits to the measured clustering statistics \citep{Hadzhiyska2020, Xu2021, AbacusHOD}.
Specifically, we use $r_{\rm vir}/r_{\rm s}$ as the halo concentration, and define the local environment for each halo as the total mass of all neighboring host halos enclosed within $3\,h^{-1}\rm Mpc$. The choice of $3\,h^{-1}\rm Mpc$ is empirical, as it yields the best fit to the target clustering statistics among four tested scales ($3, 5, 8, 10 \,h^{-1}\rm Mpc$). 

The dHOD framework implements assembly bias through the ``swapping'' technique, in which galaxies are redistributed among halos of the same mass according to the rank of a secondary property, preserving the mean occupation $\left\langle N_{\rm gal}|M \right\rangle$.
In practice, halos in the same narrow mass bin are divided into an upper and a lower percentile according to a chosen secondary property, and the average number of galaxies in the upper percentile halos is shifted by some $\delta N_{\rm gal}$, with a compensating shift in the lower percentile. 
The sign and amplitude of the shift are set by the assembly bias parameters $\mathcal{A}_{\rm cen}$ and $\mathcal{A}_{\rm sat}$, which act on centrals and satellites independently with $\mathcal{A}_{\rm cen}, \mathcal{A}_{\rm sat} \in [-1, 1]$. Positive values increase the mean occupation of halos with larger values of the secondary property, and negative values decrease it.
The default split of the halos occurs at the median of the secondary property at fixed halo mass.
However, since assembly bias may affect only a small fraction of halos \citep{Splashback}, we instead treat the split for centrals as a free parameter. 
Section~\ref{sec: fit} directly visualizes the contribution of each secondary property to assembly bias, motivating this additional flexibility. 
We fix the split at the median for satellites, as satellite assembly bias is expected to be much weaker. 

The AbacusHOD framework incorporates assembly bias by shifting the characteristic central and satellite mass scales according to each halo’s rank in the secondary property:
\begin{equation}
    \log_{10}M_{\rm cut}^{\rm mod} = \log_{10}M_{\rm cut} - \mathcal{B}_{\rm cen}(c^{\rm rank}-0.5),
\end{equation}
\begin{equation}
    \log_{10}M_{1}^{\rm mod} = \log_{10}M_{1} - \mathcal{B}_{\rm sat}(c^{\rm rank}-0.5),
\end{equation}
where $c$ denotes the chosen secondary property. This property is ranked within narrow halo mass bins and normalized to $c^{\rm rank}$, which ranges from $0$ to $1$. As a result, assembly bias parameters in the AbacusHOD framework modify the mean galaxy occupation at fixed mass directly.
We note that the sign convention for the secondary bias parameters in our implementation is opposite to that adopted in \cite{AbacusHOD}, to keep the physical implications of assembly bias consistent with the dHOD framework: specifically, a positive bias parameter
increases the occupation of halos with higher concentration or in denser local environments.
We also denote the AbacusHOD bias parameters by $\mathcal{B}$ to distinguish them from the dHOD parameters $\mathcal{A}$, as their numerical values are not directly comparable. 

All HOD models in this work are implemented with the Python package \textsc{Halotools} \citep{halotools}.

\section{Methodology} \label{sec: method}

\subsection{Clustering Statistics}

We use the projected correlation function as the summary two-point statistic throughout this work. 
As we only work with periodic boxes, the redshift space correlation function can be estimated analytically as
\begin{equation}
    \xi(r_{p}, \pi) = \frac{DD(r_{p}, \pi)}{RR(r_p, \pi)} - 1 = \frac{DD(r_p, \pi)}{N^2}\frac{V_{\rm box}}{\delta V(r_p, \pi)}-1,
\end{equation}
where $r_p$ is the transverse distance and $\pi$ is the distance along the line of sight (LOS), which we always assume to be the z-axis. 
$DD$ is the normalized pair counts between galaxies and $N$ is the total number of galaxies. Specifically, we adopt $15$ logarithmic bins along the projected separation $r_{p}$ spanning the range of $0.5\,\rm Mpc$ to $30\,\rm Mpc$, with a maximum LOS separation of $\pi_{\rm max}=60\,\rm Mpc$.

The projected correlation function is then obtained by integrating the anisotropic 2D correlation function $\xi(r_{p}, \pi)$ along the LOS,
\begin{equation}
    w_{p}\left(r_p\right)=\int_{-\pi_{\rm max}}^{\pi_{\rm max}}\xi\left(r_p,\pi\right)\ d\pi.
\end{equation}
As $w_p$ marginalizes over the LOS, it is less informative than the full-shape $\xi(r_p, \pi)$, but it is unaffected by redshift-space distortions, which introduce extra model complexity. We choose $w_p$ as a minimal data vector that could contain cosmological information. 

All correlation calculations in this work are performed with the Python package \textsc{Corrfunc} \citep{Corrfunc}. 

\subsection{Bayesian Analysis}
We fit the target $w_p$ from UniverseMachine using importance nested sampling to obtain posteriors for the parameters of the galaxy--halo connection models, as well as the model evidence, which is calculated by integrating the likelihood over the model's parameter space. 
To perform the fitting, we assume a multivariate Gaussian likelihood for $w_p$, together with an independent error $\sigma_{n}$ for the galaxy number density:
\begin{align}
    \chi^{2} &= (w_{p,\rm data}-w_{p,\rm model})^{T}\Sigma^{-1}(w_{p,\rm data}-w_{p,\rm model})  \notag \\
    &\quad + \left(\frac{n_{\rm data}-n_{\rm model}}{\sigma_{n}}\right)^2, 
\end{align}
where the covariance matrix $\Sigma$ is estimated using the jackknife method, and $\sigma_{n}=0.05n_{\rm data}$. 
As the \aemb\ boxes are relatively small, we estimate the covariance matrix with MultiDark Planck 2 (MDPL2, \citealt{MDPL2}), a $1\,h^{-1}\rm Gpc$ $N$-body simulation at Planck cosmology, whose larger volume provides a less noisy and better converged covariance estimate.

We run \um\ on MDPL2 to generate a similar galaxy catalog as our target sample, and then divide this larger-volume catalog into $5^3$ jackknife subvolumes to compute the covariance following the standard procedure:
\begin{equation}
    \Sigma_{ij} = \frac{N_{\rm jack}-1}{N_{\rm jack}}
    \sum_{k=1}^{N_{\rm jack}}
    \left(w_{p,i}^{(k)}-\overline{w}_{p,i}\right)
    \left(w_{p,j}^{(k)}-\overline{w}_{p,j}\right),
\end{equation}
where $\overline{w}_p$ is the average of all projected correlation functions measured from the $5^3$ jackknife samples. We adopt the corrections from \citet{Mohammad&Percival} and find the jackknife covariance to be broadly stable across different numbers of jackknife regions. 
To account for the smaller volume of the \aemb\ box used to generate the target clustering data, we scale the covariance by the ratio of the simulation volumes before applying it in the likelihood analysis.

We adopt nested sampling because it estimates the model evidence as a byproduct, which is essential for model comparison and for the cosmological analysis presented later in Section~\ref{sec: CEM}. Specifically, we employ \textsc{Nautilus} \citep{Nautilus}, a deep-learning-based package for Bayesian posterior and evidence estimation, with 3000 live points and a target effective sample size of 20,000. We test several choices of hyperparameters and find the estimates of model evidence to be stable across runs.

\subsection{Cosmological Evidence Modeling} \label{sec: CEM}
To obtain cosmological constraints from a relatively small number of simulations, we adopt the method of Cosmological Evidence Modeling (CEM) introduced by \cite{CEM}. The key idea is that the posterior on cosmological parameters $\mathcal{C}$, marginalized over the galaxy--halo connection parameters $\mathcal{G}$, is directly proportional to the Bayesian evidence of the galaxy--halo connection model at a given cosmology:
\begin{align}
    P(\mathcal{C}|D) &= \int P(\mathcal{C}, \mathcal{G}|D)d\mathcal{G} \\
    &\propto \int \mathcal{L}(D|\mathcal{C}, \mathcal{G})P(\mathcal{C})P(\mathcal{G})d\mathcal{G} \propto \mathcal{Z}(D|\mathcal{C}), 
\end{align}
where $D$ denotes the observed data vector, $P(\mathcal{C}|D)$ is the posterior on cosmology, $\mathcal{L}(D|\mathcal{C}, \mathcal{G})$ is the likelihood of the data at fixed cosmology and galaxy--halo connection, and $\mathcal{Z}(D|\mathcal{C})$ is the evidence for the observation $D$ given a cosmology $\mathcal{C}$.
The marginalized likelihood or Bayesian evidence $\mathcal{Z}(D|\mathcal{C})$, referred to as the cosmological evidence, can be acquired directly by fitting the target $w_p$ using nested sampling. For each galaxy--halo connection model, we repeat the fitting with halo catalogs from all simulations with diffrent cosmologies to obtain a distribution of $\mathcal{Z}(D|\mathcal{C})$, which will be used to model the cosmological posterior. The evidence for the box used to build the target galaxy catalog with \um\ is always discarded. 

After obtaining the Bayesian evidence for all cosmological simulations, we model $\mathcal{Z}(D|\mathcal{C})$ as a function of the structure growth parameter $S_8 \equiv \sigma_8(\Omega_m/0.3)^{0.5}$ alone. We restrict the dimensionality because 21 independent cosmologies are insufficient to characterize a 7-dimensional parameter space
and because the projected correlation function is most sensitive to the amplitude of matter fluctuations and the matter energy density. 
We then describe the distribution of $\mathcal{Z}(D|\mathcal{C})$ with a simple analytic form, namely a normal distribution,
\begin{equation}
    \mathcal{Z}(D|\mathcal{C}) \sim \mathcal{Z}(D|S_8) \propto \exp{\big[-(S_8-\mu)^2/2\sigma^2\big]}. 
\end{equation}
We note that the posterior $P(S_8|D)$ must satisfy the normalization condition $\int P(S_8|D)dS_8=1$, so only the shape of $\mathcal{Z}(D|S_8)$ matters. Following \cite{CEM}, we fit $\mathcal{Z}(D|S_8)$ to obtain a posterior sample of normal distributions that could fit the evidence values, and take the final cosmological posterior to be the average of all normalized Gaussians, in order to account for uncertainties in the fit.
While the Bayesian evidence is expected to be primarily a function of $S_8$, we also explore this approach for other cosmological parameters, which will be discussed in Section~\ref{sec: dis}.

This methodology reduces the expensive full joint inference over cosmology and the galaxy–halo connection to a set of independent and computationally cheaper fits of the galaxy–halo connection at each cosmology separately, allowing cosmological analysis with a limited number of simulations.

Previous CEM analyses have adopted the HOD framework for modeling the galaxy--halo connection; we verify in Appendix~\ref{app2} that CEM is also able to recover cosmological information when assuming the SHAM framework.

\section{Modeling Choices and Fits} \label{sec: fit}

\subsection{Assembly Bias in UniverseMachine}

Before constructing the empirical galaxy--halo connection models to fit target clustering, we first directly evaluate the existence and structure of assembly bias in the predictions of \um. 
Specifically, we explore whether the assembly bias signal generated by the full forward model \um\ can be reasonably represented by dependencies on a small set of parameters.
We choose halo concentration and local environment as the secondary halo properties following the conventions of formulating assembly bias, and evaluate how the galaxy occupation varies as a function of these secondary properties conditioned on halo mass.

\begin{figure*}
    \centering
    \includegraphics[width=1.9\columnwidth]{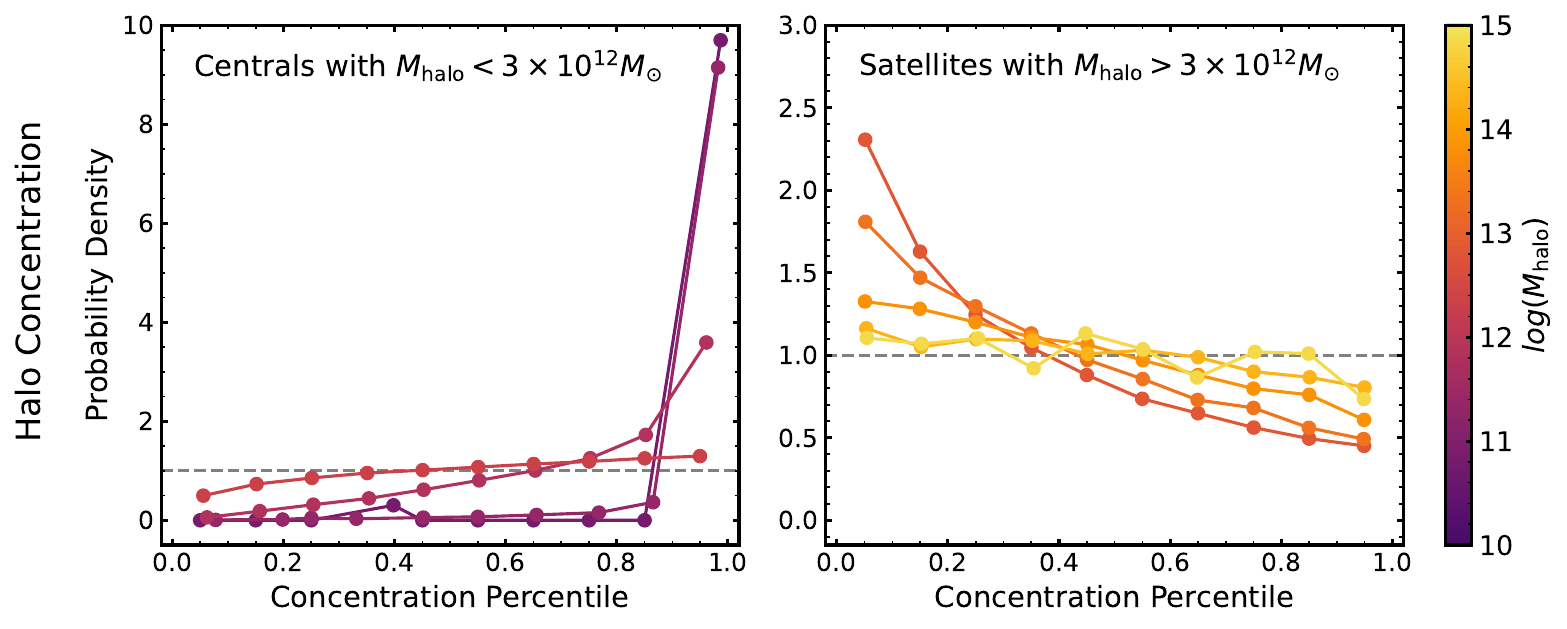}
    \includegraphics[width=1.9\columnwidth]{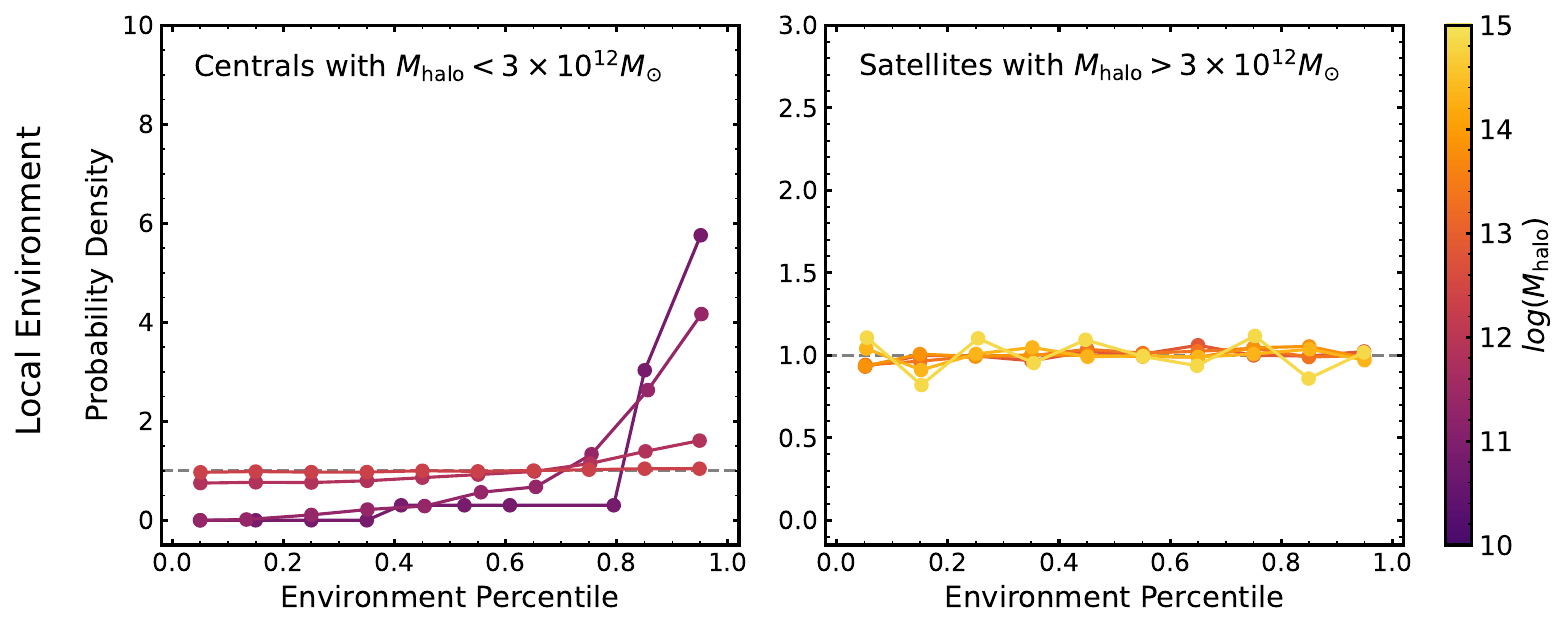}
    \caption{Probability density of galaxy occupation as a function of the percentile of a secondary halo property at fixed halo mass, $P(<c|M_{\rm halo})$, measured from the target \um\ catalog. The distribution of galaxies within halos of the same mass is characterized by quantiles of secondary halo properties. 
    The top row shows the dependence on halo concentration, and the bottom row shows the dependence on local environment, defined as the total mass of neighboring halos enclosed within $3\,h^{-1}\rm Mpc$. 
    The left panels show how the secondary halo property impacts the occupation of central galaxies in the halo mass regime of $M_{\rm halo}\lesssim 3\times10^{12}M_\odot$, which roughly corresponds to where $\left\langle N_{\rm cen} \right\rangle \lesssim 1$; the right panels show how satellite occupation varies with the secondary property in the halo mass regime of $M_{\rm halo}\gtrsim 3\times10^{12}M_\odot$. 
    The dashed gray lines at $1$ demonstrate the baseline when galaxy occupation has no additional dependence on the chosen secondary property; note that the $y$-axis limits are different between the left and right panels.
    Central occupation is positively correlated with both halo concentration and environment, while satellite occupation is negatively correlated with halo concentration only.}
    \label{fig: AB_UM}
\end{figure*}

To remove the dependence on mass and isolate the influence of secondary halo properties on the expected number of galaxies, we compute the conditional percentile $P(<c|M_{\rm halo})$, the probability of halos of mass $M_{\rm halo}$ with a secondary property below $c$, for each halo. 
Fig.~\ref{fig: AB_UM} shows the distribution of galaxy occupation as a function of the conditional percentiles.
All halos are ranked by their secondary property $c$ within narrow mass bins, and a halo of larger $P(<c|M_{\rm halo})$ has a higher value of $c$ compared to other halos of the same mass.
Specifically, we divide halos within each mass bin further into ten equally spaced bins of $P(<c|M_{\rm halo})$ ranging from $0$ to $1$, count the number of galaxies residing in halos within each percentile bin separately, and plot the probability density of the fractional contribution from galaxies in each percentile of the secondary property to the total number of galaxies in halos of such mass. A flat distribution at $1$ across all percentiles would correspond to no additional dependence on the chosen secondary property, while non-flat distributions indicate that galaxies preferentially inhabit halos with certain secondary characteristics and thus experience assembly bias.
Conditional percentiles are calculated using the \texttt{sliding\_conditional\_percentile} function from \textsc{Halotools}. 

We evaluate assembly bias for central and satellite galaxies separately. The left column of Fig.~\ref{fig: AB_UM} shows the impact of secondary properties on the occupation of centrals in the halo mass regime where $\left\langle N_{\rm cen} \right\rangle \lesssim 1$ in the HOD measured from the \um\ mock. This separation is motivated by the fact that central assembly bias can only affect halos of $M_{\rm halo}\lesssim 3\times10^{12}M_\odot$, as the mean central occupation saturates at $1$ for more massive halos, leaving no room for variation due to assembly bias. 
Halos more massive than this transition across different secondary property percentiles contribute equally to the total number of centrals, leading to flat fractional contributions. Conversely, the right column of the figure shows the impact of secondary properties on the occupation of satellites in the halo mass regime of $M_{\rm halo}\gtrsim 3\times10^{12}M_\odot$, as satellite galaxies only become abundant in this regime and thus the assembly bias signal is statistically important. 

The top row of Fig.~\ref{fig: AB_UM} demonstrates how galaxy occupation depends on the halo concentration. The left panel shows a positive correlation between central occupation and halo concentration, with a strong preference for central galaxies to reside in the highest-concentration halos, especially at lower halo masses. This is consistent with higher-concentration halos forming earlier and having deeper potential wells at fixed mass, making them more efficient at hosting a central galaxy \citep{Wechsler2002, Gao2007}. 
We suspect the strong signal in the highest concentration percentile is potentially due to a population of misclassified subhalos in the simulation, the splashback halos located outside the virial radius \citep{Diemer&Kravtsov2014}. These halos efficiently form stars at their peak mass but subsequently undergo severe mass loss, leading to low $M_{\rm vir}/M_{\rm peak}$ ratios and hence large concentrations; meanwhile, they are known to contribute significantly to halo assembly bias \citep{Splashback}. 
This observation also motivates us to include the split as a free parameter for central assembly bias in the dHOD framework as discussed in Section~\ref{sec: HOD}, since the default $50\%$ split clearly does not reflect the true assembly bias structure. We also note that this evaluation of assembly bias will not be accessible in real observations, so we allow the split to vary freely in the fit rather than fixing it to a value inferred from these plots. 
In general, the observed trend that centrals preferentially occupy higher-concentration halos at fixed mass is consistent with the clustering amplitude of \um\ being higher than the corresponding measurement in a catalog constructed from the direct tabulation of $P(N\vert M_{\rm halo})$ in \um, since higher-concentration low-mass halos tend to be more strongly clustered.

On the other hand, the right panel shows a negative correlation between satellite counts and halo concentration, indicating that satellites prefer to occupy lower-concentration host halos at fixed halo mass. This makes sense, as higher-concentration halos tend to form earlier, allowing more time for processes such as tidal stripping and dynamical friction to occur, and thus host fewer satellites \citep{vdBosch2005}. 
Meanwhile, the magnitude of concentration-based assembly bias for satellites is much smaller than for centrals, which agrees with expectations.

Similarly, the bottom row of Fig.~\ref{fig: AB_UM} shows how galaxy occupation varies with the local environment, defined as the total mass of neighboring halos enclosed within $3\,h^{-1}\rm Mpc$. We observe that central galaxies tend to occupy halos in denser environments, potentially reflecting more efficient mass assembly supported by large scale overdensities beyond single-halo scale. While the centrals preferentially occupy halos in denser environments, the trend is weaker than in the concentration case. 
In particular, the preference for halos in the uppermost percentiles of environment is significantly weaker, with central occupation increasing more gradually with the large-scale environment.
Meanwhile, satellite occupation appears unaffected by the local environment at fixed halo mass, with halos across all environment percentiles hosting approximately equal numbers of satellites, as indicated by the flat distribution in the right panel. 

To summarize, we find that the complex galaxy--halo connection physics in \um\ gives rise to a clear dependence of galaxy occupation on halo properties beyond mass.
This is reflected in the consistent variations of galaxy counts with concentration and environment (especially for centrals), suggesting both properties as reasonable and meaningful proxies to encode assembly bias in galaxy occupation, though the bias is more strongly correlated with the halo concentration than with the local environment in \um.
To recover the \um\ galaxy--halo connection and perform robust cosmological analysis with empirical galaxy--halo connection models, we need to capture the assembly bias signal as faithfully as possible to describe the clustering and to avoid biasing the cosmological inference. 

Different assembly bias implementations allow the extended HOD models to predict different shapes for the occupation distribution of Fig.~\ref{fig: AB_UM}.
The dHOD formalism produces a step function, with the split parameter setting the location of the transition, whereas the AbacusHOD formalism produces a roughly linear and more gradual variation with the conditional percentile.
However, compared to the actual measurements from \um, neither form is able to capture the real assembly bias structure. Each may be better suited to a particular case, for example, the step function to the sharper concentration dependence. The range of secondary dependence that is allowed in each model is closely connected to whether they are flexible enough to describe real galaxy populations connected to halos and subhalos, and to the cosmological results we present in Section~\ref{sec: cos}. 

\subsection{Model Comparison} \label{sec: model_comp}

\begin{figure}
    \centering    
    \includegraphics[width=0.95\columnwidth]{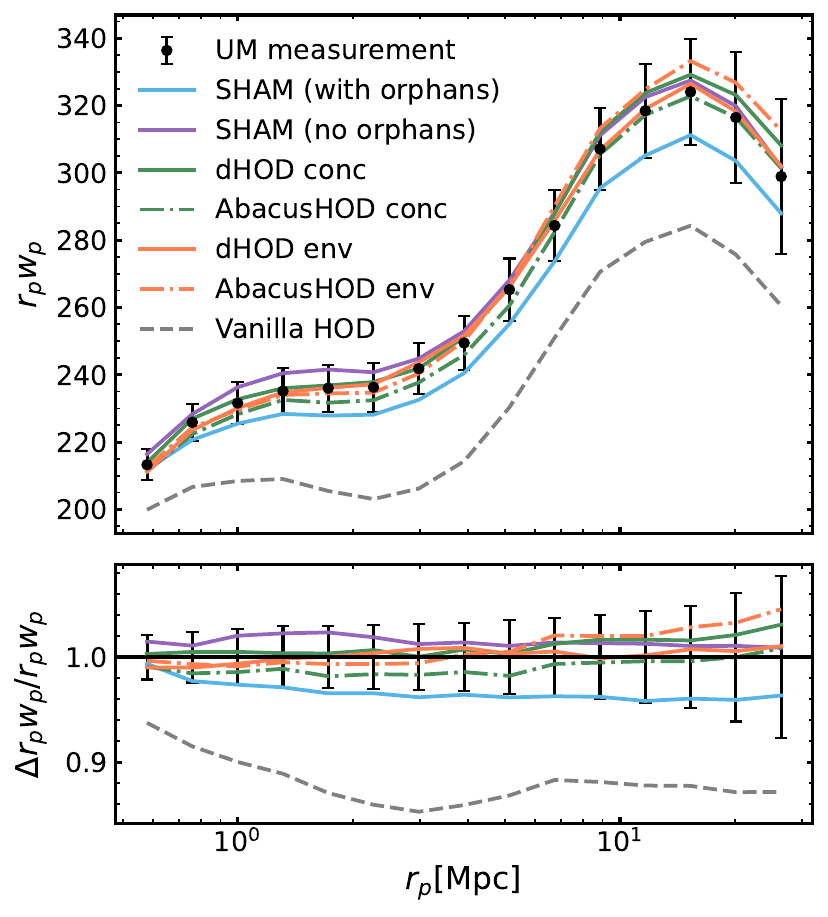}
    \caption{Fit to the \um\ $w_p$ from models introduced in Section~\ref{sec: model}, including two-parameter SHAM models built on halo catalogs with and without orphans, four HOD models with different implementations of assembly bias, and the vanilla HOD model. The lower panel shows the fractional residuals between the models and the data. All models except the vanilla HOD fit the target clustering well and yield comparable $\chi^2$ values.\footref{footnote}
    }
    \label{fig: fit}
\end{figure}

We now assess the flexibility of the SHAM and HOD models studied in this work to describe galaxy clustering properties of the \um\ catalog at the true cosmology. We fit each model to the target \um\ $w_p$ using the halo catalog of the \aemb\ box on which \um\ was built to isolate the effects of the galaxy--halo connection.
Fig.~\ref{fig: fit} presents the predicted $w_p$ from the best-fit SHAM and HOD models compared to the \um\ measurement, with error bars from the diagonal of the covariance; the lower panel shows the residuals of each model. Parameters of each model and the best-fit $\chi^2$ are summarized in Table~\ref{tab: models}. The $\chi^2$ values are consistently small for all models that fit the target clustering well, which we attribute to the models being sufficiently flexible, and to the test models and the target being built on the same simulation box and therefore sharing the same realization of cosmic variance. We thus include these values only to compare models fit to the same target.

\begin{table*}
\centering
\caption{Model parameters (Section~\ref{sec: model}), best-fit $\chi^2$ at fixed cosmology (Section~\ref{sec: fit}), and inferred cosmological constraints (Section~\ref{sec: cos}) for each empirical galaxy--halo connection model tested in this work.}
\label{tab: models}
\begin{tabular}{llcc}
\toprule
Model & Parameters & Best-fit $\chi^2$ & Inferred $S_8$ \\
\midrule
SHAM (with orphans) & $\alpha$, $s$ & 4.8 & $0.844^{+0.028}_{-0.026}$ \\
SHAM (no orphans) & $\alpha$, $s$ & 4.5 & $0.843^{+0.015}_{-0.015}$ \\
\midrule
Vanilla HOD & $M_{\rm cut}$, $M_{1}$, $\sigma$, $\alpha_{\rm HOD}$, $\kappa$ & 33.3 & $0.987^{+0.068}_{-0.059}$ \\
dHOD concentration & $+\,\mathcal{A}_{\rm cen}$, $\mathcal{A}_{\rm sat}$, $p_{\rm split, cen}$ & 4.2 & $0.879^{+0.068}_{-0.044}$ \\
AbacusHOD concentration & $+\,\mathcal{B}_{\rm cen}$, $\mathcal{B}_{\rm sat}$ & 4.0 & $0.903^{+0.070}_{-0.045}$ \\
dHOD environment & $+\,\mathcal{A}_{\rm cen}$, $\mathcal{A}_{\rm sat}$, $p_{\rm split, cen}$ & 3.5 & $0.946^{+0.126}_{-0.105}$ \\
AbacusHOD environment & $+\,\mathcal{B}_{\rm cen}$, $\mathcal{B}_{\rm sat}$ & 5.1 & $0.936^{+0.116}_{-0.084}$ \\
\bottomrule
\end{tabular}
\end{table*}

When built on the modified halo catalog that incorporates orphans using \um's galaxy disruption model, the two-parameter SHAM model that effectively adjusts satellite fraction fits the target $w_p$ well. Surprisingly, the SHAM model built on the original \textsc{Rockstar} halo catalog, which does not include orphans, is also able to fit the target $w_p$ comparably well.
This observation suggests that the SHAM parametrization introduced in Section~\ref{sec: SHAM} is able to emulate the effect of orphan modeling for the specific data vector $w_p$; this may indicate that two-point clustering of the stellar-mass-selected sample with $\log{M_{*}\geq 10.5}$ only offers modest constraining power on this low-mass galaxy population with large orphan fraction. In Appendix~\ref{app3}, we further investigate the impact of orphans for SHAM modeling by comparing the galaxy--halo connection predicted by these two implementations. We proceed to cosmological analysis using both models as they both demonstrate enough flexibility to capture the galaxy--halo connection encoded in $w_p$, raising the question of how well they marginalize over uncertainty in the galaxy--halo connection while preserving cosmological information. 

As expected, the vanilla HOD model without assembly bias fails to reproduce the clustering of \um, while all four extended HOD models with different assembly bias prescriptions provide reasonable fits to the target $w_p$. There is no significant difference in the quality of the fits between the dHOD and AbacusHOD formalisms. Both SHAM models and the four HOD models that incorporate assembly bias all yield comparable $\chi^2$ values.\footnote{We note that the error bars are strongly correlated in reality while in the figure we show only the diagonal elements of the covariance matrix, so the visual difference between the model and the data compared to the magnitudes of the error bars can be misleading. \label{footnote}} This result indicates that both empirical methods offer comparable flexibility in reproducing the clustering of galaxies in an observationally realistic mock dataset with assembly bias. 

Of course, the purpose of this modeling effort is not merely to reproduce an observed data vector, but to derive unbiased constraints on cosmology and the galaxy--halo connection; we present a comparative analysis of this capability of these models in the next section.

\section{Cosmological Results} \label{sec: cos}

In this section, we present the results of cosmological inference using CEM for each empirical model. 

\subsection{SHAM} \label{subsec: cos_sham}
\begin{figure}
    \centering
    \includegraphics[width=0.95\columnwidth]{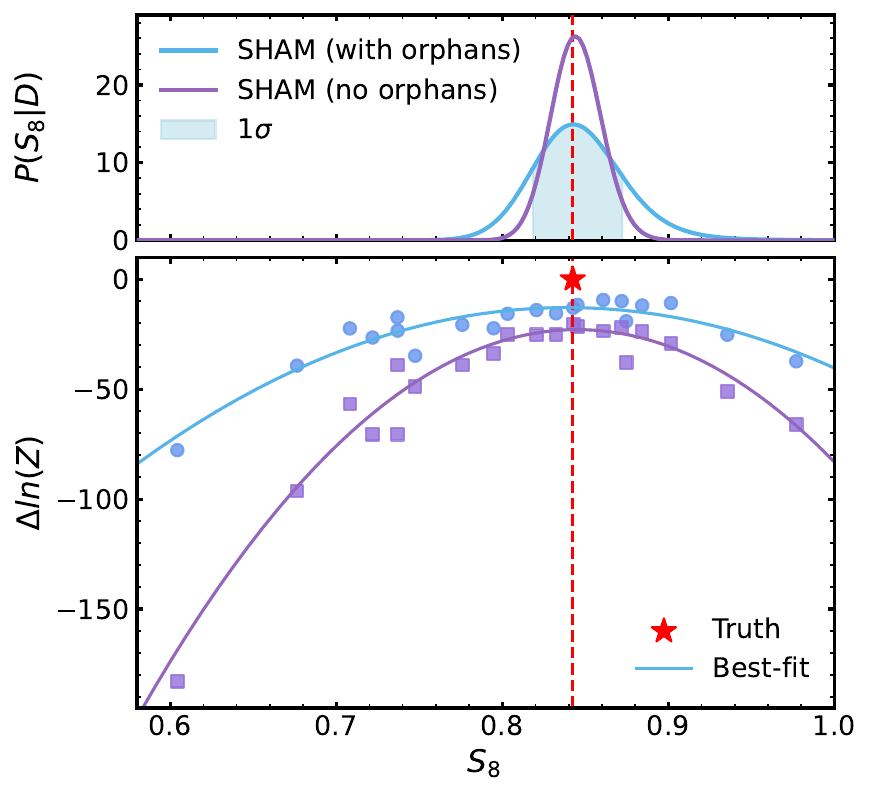}
    \caption{CEM results for the SHAM model built on halo catalogs with orphans (blue) and without orphans (purple). The lower panel plots the distribution of cosmological evidence from the \aemb\ cosmologies as a function of $S_8$, subtracting $\ln{\mathcal{Z}}$ of the correct cosmology as the reference. 
    $\Delta\ln(Z)$ from the with-orphans model are shown as blue dots, while those from the no-orphans model are plotted as purple squares and shifted down by $10$ for visual clarity, and the solid curves demonstrate the best-fit normal distributions to the evidence values for both models. 
    The upper panel shows the modeled posterior constraints on $S_8$, with the red dashed vertical line indicating the truth and the shaded region denoting the $1\sigma$ posterior interval for the with-orphans model. Both SHAM models yield unbiased constraints on $S_8$, while the constraint from the no-orphans model is tighter.}
    \label{fig: SHAM_CEM}
\end{figure}

We build the two-parameter SHAM model introduced in Section~\ref{sec: SHAM} on halo catalogs from each of the 22 \aemb\ simulations, using two halo catalogs per simulation: the standard \textsc{Rockstar} halo catalog and the modified halo catalog that incorporates orphans following \um's disruption model. This is done to assess the impact of uncertainty in orphan modeling independently. 
We fit both SHAM models to the target \um\ $w_p$ and record the Bayesian evidence $\mathcal{Z}(D|\mathcal{C})$ for all 22 cosmologies. The lower panel of Fig.~\ref{fig: SHAM_CEM} shows the values of cosmological evidence for the with-orphans model (blue) as a function of $S_8$, where $\ln{\mathcal{Z}}$ of the correct cosmology has been subtracted as a reference and is marked with a red star at zero. The values of cosmological evidence for the no-orphans model (purple) are shifted down by $10$ for better visualization. 
For both models, we find that $\ln{\mathcal{Z}}$ is maximized at the correct cosmology, and the remaining evidence values follow an approximately parabolic shape, suggesting that the choice of $S_8$ as the primary variable is reasonable and that SHAM remains sensitive to the underlying cosmology even after marginalizing over the galaxy--halo connection. 

Following the CEM methodology outlined in Section~\ref{sec: CEM}, we derive the posterior of the cosmological parameter $S_8$. For each model, the solid curve in the lower panel of Fig.~\ref{fig: SHAM_CEM} shows the best-fit Gaussian distribution to the evidence values $\mathcal{Z}(D|\mathcal{C})$, excluding that of the true cosmology.
The resulting CEM posteriors for both models are shown in the upper panel and listed in Table~\ref{tab: models}.
In both cases, we find the CEM posterior of $S_8$ to be unbiased, with the input value lying well within the $1\sigma$ interval; the shaded region shows this interval for the with-orphans model as an example.
We obtain $S_8 = 0.844 \pm 0.027$ for the with-orphans model and $S_8 = 0.843 \pm 0.015$ for the no-orphans model, while the input truth is $S_8 = 0.842$. The no-orphans model yields a tighter constraint while both remain unbiased. In general, we are able to constrain $S_8$ to within $\sim 1.8-3.2\%$ when using SHAM in the CEM analysis. 

Interestingly, the two-parameter SHAM model, whether built on the halo catalog with or without orphans, is able to infer unbiased cosmology, which complements the finding in Section~\ref{sec: model_comp} that both models can fit the target \um\ clustering well. 
This suggests that our SHAM parametrization is able to account for a range of models for subhalo stripping and still remain sensitive to cosmology, offering sufficient flexibility to absorb the complexity in both the halo catalog and the galaxy--halo connection without conflating it with cosmological information.
The different widths of the two posteriors are connected to the interpolation ranges allowed by the with-orphans and no-orphans models, which we discuss further in Appendix~\ref{app1}.

Our results demonstrate that the two-parameter SHAM model serves as a robust framework for marginalizing over galaxy--halo connection uncertainties, providing sufficient parametric flexibility to capture the essential features of the complex galaxy assembly in \um\ while remaining sensitive to cosmology. Importantly, as a modeling framework that naturally leverages more subhalo-level information from the simulations, SHAM is able to produce tight and unbiased cosmological constraints even with a relatively small number of simulations spanning a limited number of cosmologies.

\subsection{HOD} \label{sec: HOD cosmology}
We repeat the same CEM analyses for all four extended HOD formulations studied in this work: each model incorporates assembly bias through either halo concentration or local environment as the secondary dependency and is implemented using either the dHOD or AbacusHOD framework. The vanilla HOD model is also included in the cosmological analysis as a benchmark for comparison. 

For all five models, we fit them to the target \um\ clustering $w_p$ and record the Bayesian evidence $\mathcal{Z}(D|\mathcal{C})$ for all \aemb\ cosmologies. We find that $\ln{\mathcal{Z}}$ is maximized at the true cosmology for all four extended HOD models that incorporate assembly bias, while for the vanilla HOD model, the evidence peaks at a different cosmology from the input truth, indicating that the vanilla HOD model attempts to compensate for the impact of assembly bias through tweaking cosmological parameters. This behavior is linked to the results shown in Section~\ref{sec: model_comp} that all four extended HOD models are able to fit a non-trivial set of mock data from \um\ with more complicated underlying physics, while the vanilla HOD model fails the test. These results indicate the necessity of including assembly bias when modeling galaxy clustering and performing cosmological inference using HODs. 

\begin{figure}[t!]
    \centering
    \includegraphics[width=0.95\columnwidth]{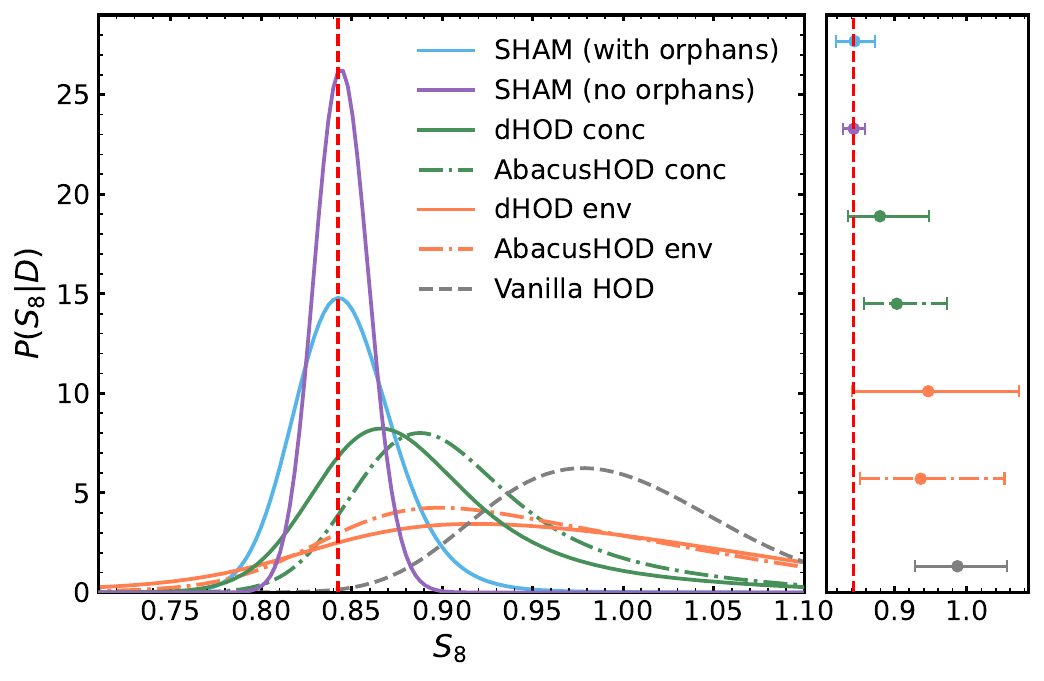} 
    \caption{Comparison of cosmological constraints on $S_8$ inferred using HOD models with different assembly bias implementations and SHAM models built on halo catalogs with and without orphans. Extended HOD models incorporating secondary dependence on the same halo property are shown in the same color, and the two linestyles denote the two implementation techniques. The right panel shows the $1\sigma$ posterior range for each model compared to the truth. All HOD models considered here yield potentially biased and weaker cosmological constraints compared to the SHAM models.}
    \label{fig: HOD_CEM}
\end{figure}

We then fit the evidence values for each model with a normal distribution and calculate the final cosmological posterior of $S_8$. 
Fig.~\ref{fig: HOD_CEM} shows the posteriors constructed using different HOD models, along with the SHAM results for comparison, with the true $S_8$ value indicated by the red dashed vertical line. Parameters of each model and the resulting constraints on $S_8$ are summarized in Table~\ref{tab: models}.

From the comparison, we first note that the vanilla HOD model yields significantly biased cosmological constraints, as indicated by the dashed gray line in Fig.~\ref{fig: HOD_CEM}. As the vanilla HOD model ignores assembly bias completely, the best-fit model predicts a projected correlation function approximately $10\%$ lower than the measurements from \um. Consequently, the vanilla model demonstrates a strong preference for larger values of $S_8$ to compensate for the higher magnitude of galaxy clustering due to assembly bias through enhanced clustering of dark matter.  


Although all assembly bias implementations considered here reproduce the clustering signal similarly well, their performance diverges when inferring cosmology. 
The extended HOD models yield potentially biased constraints on $S_8$ toward higher values, although less severely than the vanilla model, with less constraining power than the SHAM models. 
The bias is at the $\sim 1\sigma$ level, with $S_8$ constrained to $\sim6\%$ for the concentration-based implementations and $\sim11\%$ for the environment-based implementations.
The asymmetry in the posteriors is due to the fact that the high-$S_8$ tail is poorly constrained, since HOD models can fit the target clustering more easily at higher $S_8$.
HOD models tend to compensate for the enhanced galaxy clustering caused by galaxy formation physics in \um\ through adjustments to cosmological parameters, indicating that the implementations tested here do not fully capture the structure of assembly bias in the target model.

Among implementations of concentration-based biases, the constraint on $S_8$ with the AbacusHOD implementation is biased towards slightly higher values compared to the dHOD case. This difference is likely attributable to the limitations of the AbacusHOD formalism to capture \um-like assembly bias as well as the dHOD formalism. Support for this interpretation is reflected in Fig.~\ref{fig: AB_UM}, which shows that galaxies in \um\ preferentially occupy halos in the highest concentration percentile, and so the dHOD approach of shuffling galaxies with a split at large percentile is more effective in describing this signal than the AbacusHOD approach of smoothly varying galaxy occupation with halo concentration.

When implementing assembly bias based on the local environment, the dHOD and AbacusHOD formalisms generally agree on the inferred posterior of $S_8$. 
Posteriors derived with environment-based assembly bias models are wider and possibly more biased than those derived with concentration-based models; this suggests that environment-based models are less effective at capturing the true assembly bias signal in \um, and less capable of distinguishing between different underlying cosmologies.
These results offer an explicit demonstration that both the choice of secondary halo property, and the implementation technique, are of consequential importance for the capability of an HOD model to derive unbiased constraints on cosmology.

In general, we find that while the inclusion of assembly bias in HOD models appears both necessary and sufficient to reproduce the projected correlation function when assuming the correct cosmology, none of the implementations tested in this work is sufficient to fully capture the actual assembly bias signal in the target model, which results in biased cosmological constraints. At the same time, the extra freedom in the assembly bias formalism and the subhalo population leaves HOD models with weaker constraining power on cosmology than the SHAM models. This reveals a fundamental tension in assembly bias modeling for HODs: the formalism can be simultaneously too flexible and physically unrealistic---flexible enough to reproduce reasonable clustering even at incorrect cosmologies, yet not sufficiently physically grounded to recover the true assembly bias signal and cosmological information.
HOD models therefore suffer from both excessive parametric flexibility and model misspecification, which together degrade and bias the cosmological constraints they yield. This emphasizes the need to use models with more physically motivated assembly bias when pursuing cosmological analyses that extend into the one-halo regime.

\section{Higher-Order Statistics} \label{sec: acm}

To complement the comparison between the empirical models, we further evaluate their performance in predicting higher-order statistics at the true cosmology as an additional test of their ability to capture the full complexity of the galaxy–halo connection.

\subsection{Methodology \& Results}

We choose to focus on two higher-order statistics as examples, counts-in-cylinders (CiC) and multiplet clustering. 
CiC measures the number of companion galaxies in a cylindrical volume of a given projected radius and line-of-sight length centered on each galaxy (e.g. \citealt{SDSSCiC, DESICiC}). We quantify this statistic through the cumulative distribution function (CDF) for a galaxy to have fewer than $N$ such companions, $P(\leq N_{\rm CiC})$, measured in $10$ linearly spaced bins between $0$ and $9$ and $10$ logarithmically spaced bins between $10$ and $100$. Since we do not incorporate redshift space distortions in this work, we adopt a projected radius of $r_{\rm CiC}=1.5\,\rm Mpc$ and a half-length of $l_{\rm CiC}=2\,\rm Mpc$ along the LoS. 

Multiplet clustering describes how small groups of nearby galaxies, or multiplets, trace dark matter and the large-scale structure \citep{multiplet}. 
We identify multiplets by grouping galaxies within a projected separation of $1.5\,\rm Mpc$ and a LoS separation of $2\,\rm Mpc$, comparable to the cylinder used to define the CiC statistic, and measure the projected cross-correlation $w_p$ between the full galaxy sample and all multiplets containing more than one galaxy.
In other words, while CiC is sensitive to the local number density of galaxies, multiplet clustering describes the clustering patterns of these small-scale configurations with respect to the large-scale density field. 

\begin{figure*}
    \centering
    \includegraphics[width=1.9\columnwidth]{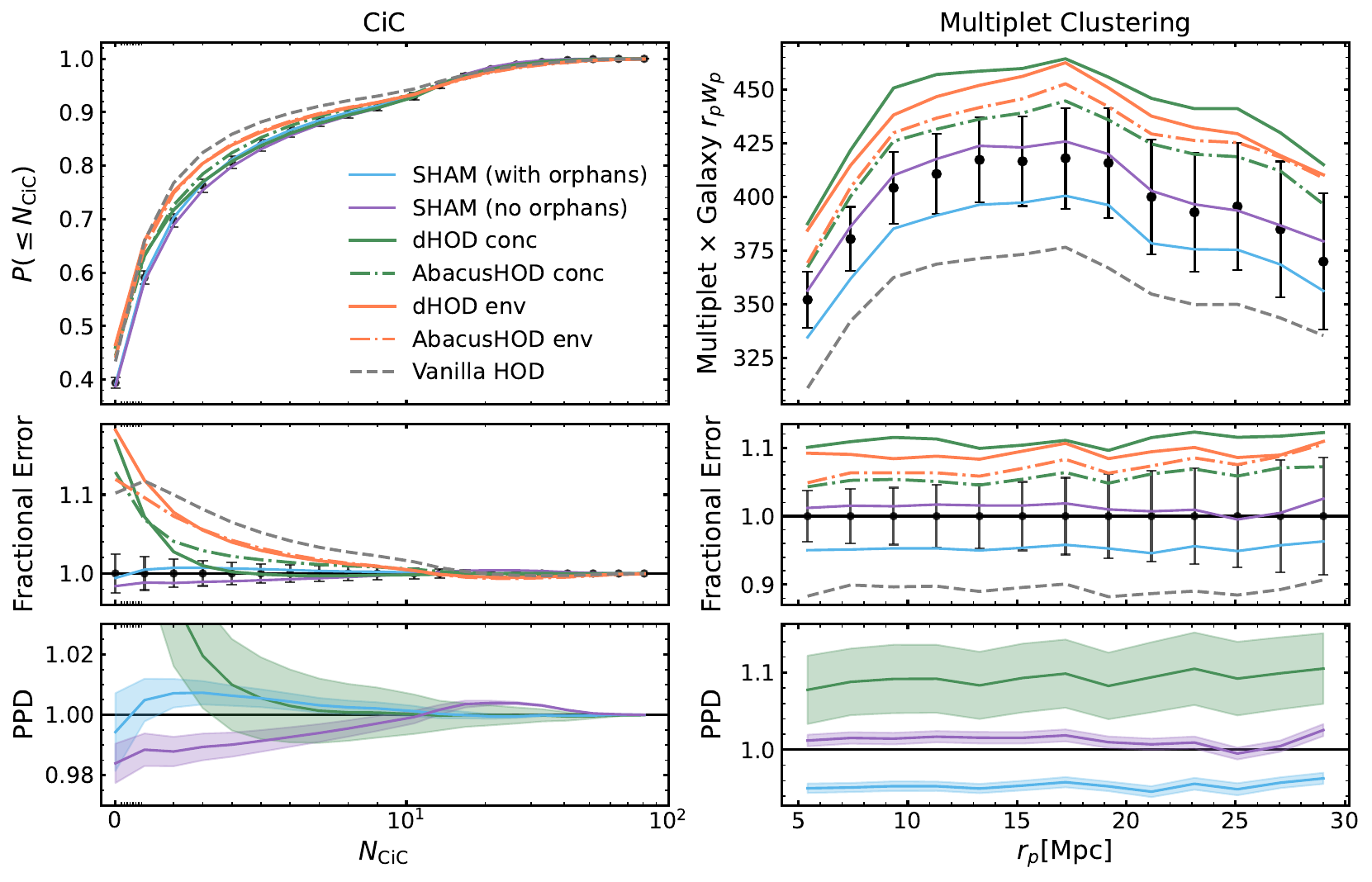}
    \caption{Higher-order statistics predictions compared to measurements from the \um\ mock (black points), for the CDF of counts-in-cylinders (CiC, left) and multiplet clustering (right). 
    For the left panels, the $x$-axis is linearly spaced between $0$ and $9$ and then logarithmically spaced between $10$ and $100$. The top panels show predictions from the best-fit galaxy--halo connection models at the true cosmology, while the middle panels show their fractional residuals with respect to \um. Jackknife error bars are included for visual guidance only.
    The bottom panels show the fractional difference between the posterior predictive distributions (PPDs) and the \um\ measurements, for the two SHAM models and the concentration-based dHOD model as an example, using samples randomly drawn from the posterior chains. Shaded regions correspond to the $1\sigma$ ranges of the PPDs, with solid curves showing their means.
    Both SHAM models predict reasonable higher-order statistics; all HOD models yield predictions that diverge more from the measurements and have wider PPDs.}
    \label{fig:acm}
\end{figure*}

For each galaxy--halo connection model, we take the best-fit parameters from the fit to the \um\ $w_p$ at the true cosmology (Section~\ref{sec: fit}) and use them to predict the higher-order statistics not used in the fit. 
The top panels of Fig.~\ref{fig:acm} show the predictions for $P(\leq N_{\rm CiC})$ and multiplet clustering from the SHAM and HOD models compared to actual \um\ measurements, while the middle panels present the fractional residuals between the predictions and the truths. 
To facilitate comparison between model predictions and the data, we include jackknife error bars for both statistics, computed in the same manner as for $w_p$. However, we note that the jackknife covariance for CiC is likely unreliable and underestimated, as we observe strong scale dependence and noisy off-diagonal terms in multiple robustness tests. The errors are also strongly correlated between bins for both statistics. We therefore show the error bars here for visual guidance only and note that robust statistical inference will require more careful covariance estimation in the future.

In order to further explore the information content of each higher-order statistic for different models, we also compute the posterior predictive distributions (PPDs), the distribution of model predictions obtained by evaluating each statistic using samples drawn from the posterior chains.
The bottom panels of Fig.~\ref{fig:acm} show the fractional difference between the PPDs and the \um\ measurements, for the two SHAM models and one example HOD model, which implements concentration-based assembly bias using the dHOD framework. The solid curves show the mean of each PPD, and the shaded regions show the $1\sigma$ intervals.

For both statistics, we observe that the best-fit predictions from the SHAM models generally agree well with the \um\ measurements, while the predictions from all HOD models deviate more significantly from the measurements. 
Moreover, the PPD of the HOD model is considerably wider than those of the SHAM models, indicating stronger parameter degeneracy in HODs that $w_p$ alone cannot resolve and explicitly demonstrating the greater parametric flexibility.

\subsection{Implications}

The ability of SHAM to reasonably predict both higher-order statistics that the model is not fit to demonstrates its robustness and its capacity to capture the essential features of the complex halo-assembly-based \um\ model, further reinforcing its potential to support unbiased cosmological inference. 


While the SHAM predictions are generally consistent with the measurements, we observe hints that SHAM models built on halo catalogs with and without orphans are intrinsically distinct given their different predictions for the higher-order statistics. In particular, the counts and clustering patterns of small galaxy configurations respond differently to orphan modeling.
The two orphan prescriptions used here represent two bracketing cases by construction: \um's disruption model likely overpredicts the orphan population, while the raw halo catalog ignores them entirely, so a realistic treatment plausibly lies between them. The higher-order statistics are consistent with this, as the measurements and their uncertainties largely overlap the range spanned by the two predictions. This highlights the need for more careful marginalization over orphan prescriptions or for more predictive subhalo tracking with well-characterized uncertainties, and motivates further investigation of subhalo tracking and higher-order statistics sensitive to it in future small-scale analyses.

The narrow PPDs of the SHAM models suggest that the galaxy--halo connection is more tightly constrained within the SHAM framework, since SHAM leverages additional information from subhalo properties and merger trees from the simulation, supporting more reliable descriptions of the galaxy--halo connection and greater sensitivity to cosmology.

On the other hand, the failure of the best-fit HOD models to reproduce these higher-order statistics again suggests that they are not sufficient to capture the full complexity in the galaxy--halo connection, such as non-Gaussianities revealed by beyond-two-point statistics. 
The discrepancies between the predicted higher-order statistics from HOD models with different secondary bias implementations indicate that these statistics can potentially discriminate between the models and provide insight on assembly bias. 
Carefully selecting and combining higher-order statistics that are sensitive to different physical aspects of the galaxy--halo connection will be important for robustly characterizing assembly bias and its impact on small-scale cosmological inference.

The wide PPD also demonstrates that these higher-order statistics can effectively break degeneracies in the HOD parameters and contain new information along directions in the galaxy--halo connection parameter space that $w_p$ alone is not sensitive to. These higher-order statistics therefore offer additional constraining power beyond the two-point statistics, motivating their inclusion in future cosmological inference. 
However, while tuning HOD models to these statistics may further constrain the models and improve agreement for specific observables, comprehensive model calibration might remain challenging given the large variety of statistics required to fully characterize non-Gaussian features in the galaxy distribution.

\section{Discussion} \label{sec: dis}

Here we test cosmological inference with projected clustering down to small scales ($0.5\,\rm Mpc$) and evaluate the empirical models against higher-order statistics. In this context, we demonstrate that SHAM captures properties of the galaxy--halo connection effectively and gains sensitivity to cosmology by leveraging predictive information from simulations about substructure and physically motivated assembly bias. 
In contrast, the excessive parametric flexibility of existing HOD models likely allows them to fit the clustering signal despite a misspecified assembly bias formalism, limiting their robustness for cosmological inference that extends to small scales.

In particular, many analyses that fit observed clustering statistics with HOD models find that including environment-based secondary bias is important. We caution, however, that the environment-based implementations tested here substantially degrade the constraining power on cosmology --- Table~\ref{tab: models} shows that they cost nearly a factor of two in the constraints on $S_8$ compared to their concentration-based counterparts. This weaker constraining power indicates a stronger degeneracy between environment-based assembly bias and $S_8$, which is itself a plausible explanation for their improved fits to observations.
We note that a change in the choice of secondary property alone can substantially affect cosmological inference, and that a better fit to clustering is not necessarily evidence that the model captures the underlying galaxy--halo connection.

We note that the analyses presented in this work use only one target model, \um, which provides a specific choice for galaxy assembly and galaxy--halo connection. \um\ has several advantages as the target model: it is one of the very few approaches that have been directly constrained by clustering data similar to that used here, it can be run quickly on large boxes, and it naturally incorporates a physically motivated assembly bias prescription. 
Although \um\ is a robustly calibrated empirical model based on a wide range of observables, it represents only one point among plausible representations of the real universe. In addition, \um\ and SHAM share the same subhalo-level approach using dark-matter-only simulations, as the core of \um\ is similar to conditional abundance matching, although with different parameters and significantly greater complexity. As a result, treating \um\ as the ground truth is not a fully independent stress test for cosmological inference with SHAM.
Nevertheless, \um\ directly ties star formation to the full assembly histories of halos through specific assumptions and flexible functional forms, and thus the secondary dependencies in the resulting mock are not necessarily reproducible by SHAM models. 
Given that \um\ is one of the few models directly tuned to a range of observational data, with a physically motivated and substantially more complex treatment of assembly bias than the simple empirical models, we regard it as a reasonable validation target.

More comprehensive stress tests of the empirical models would require a broader set of target models. These could include a broader parameter space within \um-type empirical models, semi-analytic galaxy formation models, and hydrodynamical simulations, which use fundamentally different methods to predict assembly bias, quenching, and stellar populations. 
We plan to validate our method against different suites of hydrodynamical simulations in future work, although we note that current models have difficulty simultaneously modeling sufficient volume and physics with sufficient resolution.
In addition, we plan to expand the scope of validation tests using \texttt{Diffsky} \citep{Diffstar, DSPS, Diffstarpop} to mock the physical properties of galaxy populations from various numerically expensive simulations and potentially real observations. \texttt{Diffsky}-based mocks will be able to span a wide range in the galaxy formation parameter space, supporting more comprehensive tests of the generalizability of our method to real observations.


We also note that for all CEM analyses, we model the posterior as a normal distribution and fit $\mathcal{Z}(D|\mathcal{C})$ as a function of $S_8$ alone. These choices are largely empirically motivated as they appear sufficient to describe the distribution of evidence values we measure, while they are also limited by the small number of available simulations and by sample variance from relatively small simulation volumes. We test both choices as follows.

We try fitting a skew-normal distribution to the evidence values, and find no preference for nonzero skewness. Although genuine asymmetry may exist for the HOD models (Section~\ref{sec: cos}), we are not sensitive to such an asymmetry with a limited number of cosmologies.
We then explore potential dependence on cosmological parameters other than $S_8$, such as $w$ and $H_0$, by studying $\mathcal{Z}(D|\mathcal{C})$ as a function of each individually. We find the dependence on these parameters to be much weaker and that these parameters are unable to capture the full variation in the evidence values. 
We also explore including an additional parameter beyond $S_8$ in the model, by fitting a 2D Gaussian distribution and marginalizing over the second dimension. The greater parametric flexibility of HOD models leaves their evidence values varying too weakly with cosmology to support the additional dimension; for the more constraining SHAM models, we find the $S_8$ posterior to be generally consistent with the 1D result, though slightly broader. We therefore adopt the 1D Gaussian model as a function of $S_8$ throughout the paper, noting that a more flexible form may broaden the posteriors. 
We expect that incorporating more informative data vectors would enable investigation of asymmetric posterior models and higher-dimensional cosmological inference, and even non-parametric approaches such as Gaussian process emulation \citep{CEMBOSS2}. We leave a more comprehensive exploration of these modeling choices to future work.

Here we have demonstrated the possibility of performing a robust cosmological analysis with SHAM using a small number of high-resolution simulations. While recent works have carried out cosmological inference with SHAM using the full emulator approach together with scaled simulations \citep{Contreras2021, Contreras2026}, our CEM approach directly uses high-resolution simulations, avoiding numerical errors from scaling and emulation. With increasing data quality from current and next-generation galaxy surveys, this complementary approach can support high-precision cosmological inference. In future work, we plan to validate the scaling method \citep{Angulo&White} with the high-resolution \aemb\ simulations, testing whether the scaling technique can predict reliable small-scale information required by empirical models.
We have also shown that extending the CEM framework to subhalo-based models like SHAM can advance cosmological inference beyond previous applications using HODs (e.g., \citealt{CEM, CEMBOSS1, CEMBOSS2}). In particular, our cosmological constraints are comparable to or tighter than CEM results from HODs built on a larger number of simulations and results from emulation-based methods.

Therefore, we emphasize that high-resolution simulations across cosmologies, even a limited number, offer valuable information, and the advantages of CEM combined with subhalo-based models will become increasingly valuable for precise cosmological inference. Given the lower computational cost of CEM, it is also conceivable to adopt more complex forward models like \um\ directly as the galaxy--halo connection framework, enabling more complete marginalization over uncertainties in the galaxy--halo connection against a wider range of observables and supporting more robust cosmological inference.

Going forward, we plan to apply the method to real observations. The DESI BGS sample is an immediate and well-suited target, while the method can be generalized to other samples with more careful modeling of survey-specific selection effects. In particular, with the high-density sample expected from DESI Run 2, the method could potentially extract significantly more constraining power from small scales. Here we have chosen a minimal data vector for the cosmological constraints; redshift-space multipoles would add anisotropic information, constraining the modeling of velocity space. Furthermore, incorporating higher-order statistics directly into the inference pipeline would exploit their additional constraining power on the galaxy--halo connection and cosmology, and allow us to investigate optimal combinations of statistics that capture the most information.

\section{Conclusion} \label{sec: conclusion}
In this work, we test the ability of two widely adopted families of empirical galaxy--halo connection models, subhalo abundance matching (SHAM) and the halo occupation distribution (HOD), to predict galaxy clustering and support robust cosmological inference using small-scale clustering statistics. Using the \aemb\ simulation suite, we generate a realistic stellar-mass-selected mock galaxy catalog with the more complex semi-empirical model \um, which provides a physically motivated and observationally grounded galaxy--halo connection and is a reasonable target for testing the empirical models. 

We propose a two-parameter SHAM model that allows flexible dependence on when each halo reaches its peak mass, through a halo proxy parameterization that smoothly interpolates between ranking by $M_{\rm peak}$ and ranking by $v_{\rm vir}(a_{M_{\rm peak}})$, effectively adjusting the satellite fraction. We build this model on halo catalogs both with and without orphans.
On the HOD side, we focus on a ``vanilla'' five-parameter HOD model where occupation depends only on host mass, and four extended HOD models that implement assembly bias using different secondary halo properties in different ways (detailed in Section~\ref{sec: model}). We directly investigate assembly bias in the \um\ mock and test whether our empirical models can reproduce the projected clustering $w_p$ of this mock down to small scales ($\sim 0.5\,\rm Mpc$) at the true cosmology. 

Our primary focus is to determine whether fitting these models to the target clustering statistics over the range of cosmologies sampled by \aemb\ can lead to unbiased parameter recovery. In particular, using cosmological evidence modeling, we construct constraints on the structure growth parameter $S_8$ using evidence values obtained from sampling at each different cosmology. 
Finally, we predict two higher-order statistics, counts-in-cylinders and multiplet clustering, using best-fit parameters from the fit to $w_p$ for each model, and compare the predictions to actual \um\ measurements.  

Our results show that:
\begin{itemize}
  \item Assembly bias in \um\ is more strongly correlated with the halo concentration than with the local environment (characterized by the total mass of neighboring halos enclosed within $3\,h^{-1}\rm Mpc$), and appears to be more prominent for central galaxies than for satellites (Fig.~\ref{fig: AB_UM}). 
  None of the HOD assembly bias formalisms tested in this work can capture the structure of assembly bias in \um.
  
  \item At fixed cosmology, the two-parameter SHAM model is able to reproduce \um's projected two-point correlation function $w_p$ both with the raw \textsc{Rockstar} halo catalog and with a modified halo catalog that incorporates orphans (Fig.~\ref{fig: fit}), suggesting that our SHAM parameterization can potentially encapsulate the still-uncertain effects of subhalo tidal stripping for the data vector considered here. 
  
  \item At fixed cosmology, the four extended HOD models with assembly bias are all flexible enough to reproduce \um's projected two-point correlation function $w_p$, while the vanilla HOD model does not have this capability (Fig.~\ref{fig: fit}).
  
  \item The two-parameter SHAM model is able to offer tight, unbiased cosmological constraints on the structure growth parameter $S_8$ even with simulations of relatively small volumes that span a limited number of cosmologies (Fig.~\ref{fig: SHAM_CEM}). This result holds for two assumptions about orphan modeling that likely bracket the true subhalo tidal stripping, and suggests that SHAM is a reliable galaxy--halo connection framework to support robust cosmological inference on small scales.

  \item While HOD models that incorporate assembly bias are able to reproduce $w_p$ at fixed cosmology, they result in potentially biased cosmological constraints that are weaker than those of SHAM models and sensitive to how assembly bias is implemented. 
  HOD models attempt to compensate for complexity in the galaxy--halo connection through adjustments in cosmological parameters; the combination of excessive parametric flexibility and model misspecification degrades and biases cosmological constraints from HODs (Fig.~\ref{fig: HOD_CEM}).
  
  \item The SHAM models again demonstrate robustness by successfully reproducing higher-order statistics (counts-in-cylinders and multiplet clustering) that are not used in the fit, whereas HOD models fail this validation test, revealing limitations in their ability to capture higher-order aspects of the galaxy--halo connection with sufficient accuracy (Fig.~\ref{fig:acm}). This also indicates that including beyond-two-point statistics can further support cosmological analyses on small scales by distinguishing between models with different assembly bias treatments.

\end{itemize}

In summary, projected two-point clustering alone can robustly constrain cosmology when using predictive subhalo-level information, while more parametric models such as HODs provide comparatively weaker and potentially biased constraints, motivating the use of subhalo-based models for cosmological inference on small scales.

\section*{Acknowledgments}
This work received support from the Kavli Institute for Particle Astrophysics and Cosmology at Stanford and SLAC, and from the U.S. Department of Energy under contract number DE-AC02-76SF00515 to SLAC National Accelerator Laboratory. HW received support from the Stanford Graduate Fellowship. Work done at Argonne was supported under the DOE contract DE-AC02-06CH11357.

We thank Joe DeRose, Yao-Yuan Mao, Shanthala Ashwath, and Shawn McLaughlin for contributions to running the \aemb\ simulation suite. We thank the KIPAC Galaxy Formation and Cosmology group for useful discussions, Joe DeRose for helpful comments on a draft, Tara Dacunha for helpful advice about the visualizations in Figure 1, and Yao-Yuan Mao for useful discussions about the two-parameter SHAM model.

This research made use of computational resources on the Sherlock cluster at Stanford and the S3DF System at SLAC. S3DF is a shared High-Performance Computing facility that supports the scientific and data-intensive computing needs of all experimental facilities and programs of the SLAC National Accelerator Laboratory. SLAC is operated by Stanford University for the U.S. Department of Energy’s Office of Science. We thank Stanford University and the Stanford Research Computing Center, as well as the SLAC computational team, for computational resources and support that contributed to these results.

\bibliography{main.bib}

@ARTICLE{BWS2002,
       author = {{Bullock}, James S. and {Wechsler}, Risa H. and {Somerville}, Rachel S.},
        title = "{Galaxy halo occupation at high redshift}",
      journal = {\mnras},
         year = 2002,
        month = jan,
       volume = {329},
       number = {1},
        pages = {246-256},
          doi = {10.1046/j.1365-8711.2002.04959.x},
archivePrefix = {arXiv},
       eprint = {astro-ph/0106293},
 primaryClass = {astro-ph},
       adsurl = {https://ui.adsabs.harvard.edu/abs/2002MNRAS.329..246B}
}

@PHDTHESIS{Wechsler2001,
       author = {{Wechsler}, Risa H.},
        title = "{Dark halo merging and galaxy formation}",
       school = {University of California, Santa Cruz},
         year = 2001,
        month = jan,
       adsurl = {https://ui.adsabs.harvard.edu/abs/2001PhDT.........7W}
}

@ARTICLE{Mao2018,
       author = {{Mao}, Yao-Yuan and {Zentner}, Andrew R. and {Wechsler}, Risa H.},
        title = "{Beyond assembly bias: exploring secondary halo biases for cluster-size haloes}",
      journal = {\mnras},
         year = 2018,
        month = mar,
       volume = {474},
       number = {4},
        pages = {5143-5157},
          doi = {10.1093/mnras/stx3111},
archivePrefix = {arXiv},
       eprint = {1705.03888},
 primaryClass = {astro-ph.CO},
       adsurl = {https://ui.adsabs.harvard.edu/abs/2018MNRAS.474.5143M}
}

@ARTICLE{Gao2005,
       author = {{Gao}, Liang and {Springel}, Volker and {White}, Simon D.~M.},
        title = "{The age dependence of halo clustering}",
      journal = {\mnras},
         year = 2005,
        month = oct,
       volume = {363},
       number = {1},
        pages = {L66-L70},
          doi = {10.1111/j.1745-3933.2005.00084.x},
archivePrefix = {arXiv},
       eprint = {astro-ph/0506510},
 primaryClass = {astro-ph},
       adsurl = {https://ui.adsabs.harvard.edu/abs/2005MNRAS.363L..66G}
}

@ARTICLE{DESI1,
       author = {{DESI Collaboration} and {Aghamousa}, Amir and {Aguilar}, Jessica and {Ahlen}, Steve and {Alam}, Shadab and {Allen}, Lori E. and {Allende Prieto}, Carlos and {Annis}, James and {Bailey}, Stephen and {Balland}, Christophe and {Ballester}, Otger and {Baltay}, Charles and {Beaufore}, Lucas and {Bebek}, Chris and {Beers}, Timothy C. and {Bell}, Eric F. and {Bernal}, Jos{\'e} Luis and {Besuner}, Robert and {Beutler}, Florian and {Blake}, Chris and {Bleuler}, Hannes and {Blomqvist}, Michael and {Blum}, Robert and {Bolton}, Adam S. and {Briceno}, Cesar and {Brooks}, David and {Brownstein}, Joel R. and {Buckley-Geer}, Elizabeth and {Burden}, Angela and {Burtin}, Etienne and {Busca}, Nicolas G. and {Cahn}, Robert N. and {Cai}, Yan-Chuan and {Cardiel-Sas}, Laia and {Carlberg}, Raymond G. and {Carton}, Pierre-Henri and {Casas}, Ricard and {Castander}, Francisco J. and {Cervantes-Cota}, Jorge L. and {Claybaugh}, Todd M. and {Close}, Madeline and {Coker}, Carl T. and {Cole}, Shaun and {Comparat}, Johan and {Cooper}, Andrew P. and {Cousinou}, M. -C. and {Crocce}, Martin and {Cuby}, Jean-Gabriel and {Cunningham}, Daniel P. and {Davis}, Tamara M. and {Dawson}, Kyle S. and {de la Macorra}, Axel and {De Vicente}, Juan and {Delubac}, Timoth{\'e}e and {Derwent}, Mark and {Dey}, Arjun and {Dhungana}, Govinda and {Ding}, Zhejie and {Doel}, Peter and {Duan}, Yutong T. and {Ealet}, Anne and {Edelstein}, Jerry and {Eftekharzadeh}, Sarah and {Eisenstein}, Daniel J. and {Elliott}, Ann and {Escoffier}, St{\'e}phanie and {Evatt}, Matthew and {Fagrelius}, Parker and {Fan}, Xiaohui and {Fanning}, Kevin and {Farahi}, Arya and {Farihi}, Jay and {Favole}, Ginevra and {Feng}, Yu and {Fernandez}, Enrique and {Findlay}, Joseph R. and {Finkbeiner}, Douglas P. and {Fitzpatrick}, Michael J. and {Flaugher}, Brenna and {Flender}, Samuel and {Font-Ribera}, Andreu and {Forero-Romero}, Jaime E. and {Fosalba}, Pablo and {Frenk}, Carlos S. and {Fumagalli}, Michele and {Gaensicke}, Boris T. and {Gallo}, Giuseppe and {Garcia-Bellido}, Juan and {Gaztanaga}, Enrique and {Pietro Gentile Fusillo}, Nicola and {Gerard}, Terry and {Gershkovich}, Irena and {Giannantonio}, Tommaso and {Gillet}, Denis and {Gonzalez-de-Rivera}, Guillermo and {Gonzalez-Perez}, Violeta and {Gott}, Shelby and {Graur}, Or and {Gutierrez}, Gaston and {Guy}, Julien and {Habib}, Salman and {Heetderks}, Henry and {Heetderks}, Ian and {Heitmann}, Katrin and {Hellwing}, Wojciech A. and {Herrera}, David A. and {Ho}, Shirley and {Holland}, Stephen and {Honscheid}, Klaus and {Huff}, Eric and {Hutchinson}, Timothy A. and {Huterer}, Dragan and {Hwang}, Ho Seong and {Illa Laguna}, Joseph Maria and {Ishikawa}, Yuzo and {Jacobs}, Dianna and {Jeffrey}, Niall and {Jelinsky}, Patrick and {Jennings}, Elise and {Jiang}, Linhua and {Jimenez}, Jorge and {Johnson}, Jennifer and {Joyce}, Richard and {Jullo}, Eric and {Juneau}, St{\'e}phanie and {Kama}, Sami and {Karcher}, Armin and {Karkar}, Sonia and {Kehoe}, Robert and {Kennamer}, Noble and {Kent}, Stephen and {Kilbinger}, Martin and {Kim}, Alex G. and {Kirkby}, David and {Kisner}, Theodore and {Kitanidis}, Ellie and {Kneib}, Jean-Paul and {Koposov}, Sergey and {Kovacs}, Eve and {Koyama}, Kazuya and {Kremin}, Anthony and {Kron}, Richard and {Kronig}, Luzius and {Kueter-Young}, Andrea and {Lacey}, Cedric G. and {Lafever}, Robin and {Lahav}, Ofer and {Lambert}, Andrew and {Lampton}, Michael and {Landriau}, Martin and {Lang}, Dustin and {Lauer}, Tod R. and {Le Goff}, Jean-Marc and {Le Guillou}, Laurent and {Le Van Suu}, Auguste and {Lee}, Jae Hyeon and {Lee}, Su-Jeong and {Leitner}, Daniela and {Lesser}, Michael and {Levi}, Michael E. and {L'Huillier}, Benjamin and {Li}, Baojiu and {Liang}, Ming and {Lin}, Huan and {Linder}, Eric and {Loebman}, Sarah R. and {Luki{\'c}}, Zarija and {Ma}, Jun and {MacCrann}, Niall and {Magneville}, Christophe and {Makarem}, Laleh and {Manera}, Marc and {Manser}, Christopher J. and {Marshall}, Robert and {Martini}, Paul and {Massey}, Richard and {Matheson}, Thomas and {McCauley}, Jeremy and {McDonald}, Patrick and {McGreer}, Ian D. and {Meisner}, Aaron and {Metcalfe}, Nigel and {Miller}, Timothy N. and {Miquel}, Ramon and {Moustakas}, John and {Myers}, Adam and {Naik}, Milind and {Newman}, Jeffrey A. and {Nichol}, Robert C. and {Nicola}, Andrina and {Nicolati da Costa}, Luiz and {Nie}, Jundan and {Niz}, Gustavo and {Norberg}, Peder and {Nord}, Brian and {Norman}, Dara and {Nugent}, Peter and {O'Brien}, Thomas and {Oh}, Minji and {Olsen}, Knut A.~G. and {Padilla}, Cristobal and {Padmanabhan}, Hamsa and {Padmanabhan}, Nikhil and {Palanque-Delabrouille}, Nathalie and {Palmese}, Antonella and {Pappalardo}, Daniel and {P{\^a}ris}, Isabelle and {Park}, Changbom and {Patej}, Anna and {Peacock}, John A. and {Peiris}, Hiranya V. and {Peng}, Xiyan and {Percival}, Will J. and {Perruchot}, Sandrine and {Pieri}, Matthew M. and {Pogge}, Richard and {Pollack}, Jennifer E. and {Poppett}, Claire and {Prada}, Francisco and {Prakash}, Abhishek and {Probst}, Ronald G. and {Rabinowitz}, David and {Raichoor}, Anand and {Ree}, Chang Hee and {Refregier}, Alexandre and {Regal}, Xavier and {Reid}, Beth and {Reil}, Kevin and {Rezaie}, Mehdi and {Rockosi}, Constance M. and {Roe}, Natalie and {Ronayette}, Samuel and {Roodman}, Aaron and {Ross}, Ashley J. and {Ross}, Nicholas P. and {Rossi}, Graziano and {Rozo}, Eduardo and {Ruhlmann-Kleider}, Vanina and {Rykoff}, Eli S. and {Sabiu}, Cristiano and {Samushia}, Lado and {Sanchez}, Eusebio and {Sanchez}, Javier and {Schlegel}, David J. and {Schneider}, Michael and {Schubnell}, Michael and {Secroun}, Aur{\'e}lia and {Seljak}, Uros and {Seo}, Hee-Jong and {Serrano}, Santiago and {Shafieloo}, Arman and {Shan}, Huanyuan and {Sharples}, Ray and {Sholl}, Michael J. and {Shourt}, William V. and {Silber}, Joseph H. and {Silva}, David R. and {Sirk}, Martin M. and {Slosar}, Anze and {Smith}, Alex and {Smoot}, George F. and {Som}, Debopam and {Song}, Yong-Seon and {Sprayberry}, David and {Staten}, Ryan and {Stefanik}, Andy and {Tarle}, Gregory and {Sien Tie}, Suk and {Tinker}, Jeremy L. and {Tojeiro}, Rita and {Valdes}, Francisco and {Valenzuela}, Octavio and {Valluri}, Monica and {Vargas-Magana}, Mariana and {Verde}, Licia and {Walker}, Alistair R. and {Wang}, Jiali and {Wang}, Yuting and {Weaver}, Benjamin A. and {Weaverdyck}, Curtis and {Wechsler}, Risa H. and {Weinberg}, David H. and {White}, Martin and {Yang}, Qian and {Yeche}, Christophe and {Zhang}, Tianmeng and {Zhao}, Gong-Bo and {Zheng}, Yi and {Zhou}, Xu and {Zhou}, Zhimin and {Zhu}, Yaling and {Zou}, Hu and {Zu}, Ying},
        title = "{The DESI Experiment Part I: Science,Targeting, and Survey Design}",
      journal = {arXiv e-prints},
         year = 2016,
        month = oct,
          eid = {arXiv:1611.00036},
        pages = {arXiv:1611.00036},
          doi = {10.48550/arXiv.1611.00036},
archivePrefix = {arXiv},
       eprint = {1611.00036},
 primaryClass = {astro-ph.IM},
       adsurl = {https://ui.adsabs.harvard.edu/abs/2016arXiv161100036D}
}

@ARTICLE{DESI2,
       author = {{DESI Collaboration} and {Aghamousa}, Amir and {Aguilar}, Jessica and {Ahlen}, Steve and {Alam}, Shadab and {Allen}, Lori E. and {Allende Prieto}, Carlos and {Annis}, James and {Bailey}, Stephen and {Balland}, Christophe and {Ballester}, Otger and {Baltay}, Charles and {Beaufore}, Lucas and {Bebek}, Chris and {Beers}, Timothy C. and {Bell}, Eric F. and {Bernal}, Jos{\'e} Luis and {Besuner}, Robert and {Beutler}, Florian and {Blake}, Chris and {Bleuler}, Hannes and {Blomqvist}, Michael and {Blum}, Robert and {Bolton}, Adam S. and {Briceno}, Cesar and {Brooks}, David and {Brownstein}, Joel R. and {Buckley-Geer}, Elizabeth and {Burden}, Angela and {Burtin}, Etienne and {Busca}, Nicolas G. and {Cahn}, Robert N. and {Cai}, Yan-Chuan and {Cardiel-Sas}, Laia and {Carlberg}, Raymond G. and {Carton}, Pierre-Henri and {Casas}, Ricard and {Castander}, Francisco J. and {Cervantes-Cota}, Jorge L. and {Claybaugh}, Todd M. and {Close}, Madeline and {Coker}, Carl T. and {Cole}, Shaun and {Comparat}, Johan and {Cooper}, Andrew P. and {Cousinou}, M. -C. and {Crocce}, Martin and {Cuby}, Jean-Gabriel and {Cunningham}, Daniel P. and {Davis}, Tamara M. and {Dawson}, Kyle S. and {de la Macorra}, Axel and {De Vicente}, Juan and {Delubac}, Timoth{\'e}e and {Derwent}, Mark and {Dey}, Arjun and {Dhungana}, Govinda and {Ding}, Zhejie and {Doel}, Peter and {Duan}, Yutong T. and {Ealet}, Anne and {Edelstein}, Jerry and {Eftekharzadeh}, Sarah and {Eisenstein}, Daniel J. and {Elliott}, Ann and {Escoffier}, St{\'e}phanie and {Evatt}, Matthew and {Fagrelius}, Parker and {Fan}, Xiaohui and {Fanning}, Kevin and {Farahi}, Arya and {Farihi}, Jay and {Favole}, Ginevra and {Feng}, Yu and {Fernandez}, Enrique and {Findlay}, Joseph R. and {Finkbeiner}, Douglas P. and {Fitzpatrick}, Michael J. and {Flaugher}, Brenna and {Flender}, Samuel and {Font-Ribera}, Andreu and {Forero-Romero}, Jaime E. and {Fosalba}, Pablo and {Frenk}, Carlos S. and {Fumagalli}, Michele and {Gaensicke}, Boris T. and {Gallo}, Giuseppe and {Garcia-Bellido}, Juan and {Gaztanaga}, Enrique and {Pietro Gentile Fusillo}, Nicola and {Gerard}, Terry and {Gershkovich}, Irena and {Giannantonio}, Tommaso and {Gillet}, Denis and {Gonzalez-de-Rivera}, Guillermo and {Gonzalez-Perez}, Violeta and {Gott}, Shelby and {Graur}, Or and {Gutierrez}, Gaston and {Guy}, Julien and {Habib}, Salman and {Heetderks}, Henry and {Heetderks}, Ian and {Heitmann}, Katrin and {Hellwing}, Wojciech A. and {Herrera}, David A. and {Ho}, Shirley and {Holland}, Stephen and {Honscheid}, Klaus and {Huff}, Eric and {Hutchinson}, Timothy A. and {Huterer}, Dragan and {Hwang}, Ho Seong and {Illa Laguna}, Joseph Maria and {Ishikawa}, Yuzo and {Jacobs}, Dianna and {Jeffrey}, Niall and {Jelinsky}, Patrick and {Jennings}, Elise and {Jiang}, Linhua and {Jimenez}, Jorge and {Johnson}, Jennifer and {Joyce}, Richard and {Jullo}, Eric and {Juneau}, St{\'e}phanie and {Kama}, Sami and {Karcher}, Armin and {Karkar}, Sonia and {Kehoe}, Robert and {Kennamer}, Noble and {Kent}, Stephen and {Kilbinger}, Martin and {Kim}, Alex G. and {Kirkby}, David and {Kisner}, Theodore and {Kitanidis}, Ellie and {Kneib}, Jean-Paul and {Koposov}, Sergey and {Kovacs}, Eve and {Koyama}, Kazuya and {Kremin}, Anthony and {Kron}, Richard and {Kronig}, Luzius and {Kueter-Young}, Andrea and {Lacey}, Cedric G. and {Lafever}, Robin and {Lahav}, Ofer and {Lambert}, Andrew and {Lampton}, Michael and {Landriau}, Martin and {Lang}, Dustin and {Lauer}, Tod R. and {Le Goff}, Jean-Marc and {Le Guillou}, Laurent and {Le Van Suu}, Auguste and {Lee}, Jae Hyeon and {Lee}, Su-Jeong and {Leitner}, Daniela and {Lesser}, Michael and {Levi}, Michael E. and {L'Huillier}, Benjamin and {Li}, Baojiu and {Liang}, Ming and {Lin}, Huan and {Linder}, Eric and {Loebman}, Sarah R. and {Luki{\'c}}, Zarija and {Ma}, Jun and {MacCrann}, Niall and {Magneville}, Christophe and {Makarem}, Laleh and {Manera}, Marc and {Manser}, Christopher J. and {Marshall}, Robert and {Martini}, Paul and {Massey}, Richard and {Matheson}, Thomas and {McCauley}, Jeremy and {McDonald}, Patrick and {McGreer}, Ian D. and {Meisner}, Aaron and {Metcalfe}, Nigel and {Miller}, Timothy N. and {Miquel}, Ramon and {Moustakas}, John and {Myers}, Adam and {Naik}, Milind and {Newman}, Jeffrey A. and {Nichol}, Robert C. and {Nicola}, Andrina and {Nicolati da Costa}, Luiz and {Nie}, Jundan and {Niz}, Gustavo and {Norberg}, Peder and {Nord}, Brian and {Norman}, Dara and {Nugent}, Peter and {O'Brien}, Thomas and {Oh}, Minji and {Olsen}, Knut A.~G. and {Padilla}, Cristobal and {Padmanabhan}, Hamsa and {Padmanabhan}, Nikhil and {Palanque-Delabrouille}, Nathalie and {Palmese}, Antonella and {Pappalardo}, Daniel and {P{\^a}ris}, Isabelle and {Park}, Changbom and {Patej}, Anna and {Peacock}, John A. and {Peiris}, Hiranya V. and {Peng}, Xiyan and {Percival}, Will J. and {Perruchot}, Sandrine and {Pieri}, Matthew M. and {Pogge}, Richard and {Pollack}, Jennifer E. and {Poppett}, Claire and {Prada}, Francisco and {Prakash}, Abhishek and {Probst}, Ronald G. and {Rabinowitz}, David and {Raichoor}, Anand and {Ree}, Chang Hee and {Refregier}, Alexandre and {Regal}, Xavier and {Reid}, Beth and {Reil}, Kevin and {Rezaie}, Mehdi and {Rockosi}, Constance M. and {Roe}, Natalie and {Ronayette}, Samuel and {Roodman}, Aaron and {Ross}, Ashley J. and {Ross}, Nicholas P. and {Rossi}, Graziano and {Rozo}, Eduardo and {Ruhlmann-Kleider}, Vanina and {Rykoff}, Eli S. and {Sabiu}, Cristiano and {Samushia}, Lado and {Sanchez}, Eusebio and {Sanchez}, Javier and {Schlegel}, David J. and {Schneider}, Michael and {Schubnell}, Michael and {Secroun}, Aur{\'e}lia and {Seljak}, Uros and {Seo}, Hee-Jong and {Serrano}, Santiago and {Shafieloo}, Arman and {Shan}, Huanyuan and {Sharples}, Ray and {Sholl}, Michael J. and {Shourt}, William V. and {Silber}, Joseph H. and {Silva}, David R. and {Sirk}, Martin M. and {Slosar}, Anze and {Smith}, Alex and {Smoot}, George F. and {Som}, Debopam and {Song}, Yong-Seon and {Sprayberry}, David and {Staten}, Ryan and {Stefanik}, Andy and {Tarle}, Gregory and {Sien Tie}, Suk and {Tinker}, Jeremy L. and {Tojeiro}, Rita and {Valdes}, Francisco and {Valenzuela}, Octavio and {Valluri}, Monica and {Vargas-Magana}, Mariana and {Verde}, Licia and {Walker}, Alistair R. and {Wang}, Jiali and {Wang}, Yuting and {Weaver}, Benjamin A. and {Weaverdyck}, Curtis and {Wechsler}, Risa H. and {Weinberg}, David H. and {White}, Martin and {Yang}, Qian and {Yeche}, Christophe and {Zhang}, Tianmeng and {Zhao}, Gong-Bo and {Zheng}, Yi and {Zhou}, Xu and {Zhou}, Zhimin and {Zhu}, Yaling and {Zou}, Hu and {Zu}, Ying},
        title = "{The DESI Experiment Part II: Instrument Design}",
      journal = {arXiv e-prints},
         year = 2016,
        month = oct,
          eid = {arXiv:1611.00037},
        pages = {arXiv:1611.00037},
          doi = {10.48550/arXiv.1611.00037},
archivePrefix = {arXiv},
       eprint = {1611.00037},
 primaryClass = {astro-ph.IM},
       adsurl = {https://ui.adsabs.harvard.edu/abs/2016arXiv161100037D}
}

@ARTICLE{Euclid,
       author = {{Euclid Collaboration} and {Mellier}, Y. and {Abdurro'uf} and {Acevedo Barroso}, J.~A. and {Ach{\'u}carro}, A. and {Adamek}, J. and {Adam}, R. and {Addison}, G.~E. and {Aghanim}, N. and {Aguena}, M. and {Ajani}, V. and {Akrami}, Y. and {Al-Bahlawan}, A. and {Alavi}, A. and {Albuquerque}, I.~S. and {Alestas}, G. and {Alguero}, G. and {Allaoui}, A. and {Allen}, S.~W. and {Allevato}, V. and {Alonso-Tetilla}, A.~V. and {Altieri}, B. and {Alvarez-Candal}, A. and {Alvi}, S. and {Amara}, A. and {Amendola}, L. and {Amiaux}, J. and {Andika}, I.~T. and {Andreon}, S. and {Andrews}, A. and {Angora}, G. and {Angulo}, R.~E. and {Annibali}, F. and {Anselmi}, A. and {Anselmi}, S. and {Arcari}, S. and {Archidiacono}, M. and {Aric{\`o}}, G. and {Arnaud}, M. and {Arnouts}, S. and {Asgari}, M. and {Asorey}, J. and {Atayde}, L. and {Atek}, H. and {Atrio-Barandela}, F. and {Aubert}, M. and {Aubourg}, E. and {Auphan}, T. and {Auricchio}, N. and {Aussel}, B. and {Aussel}, H. and {Avelino}, P.~P. and {Avgoustidis}, A. and {Avila}, S. and {Awan}, S. and {Azzollini}, R. and {Baccigalupi}, C. and {Bachelet}, E. and {Bacon}, D. and {Baes}, M. and {Bagley}, M.~B. and {Bahr-Kalus}, B. and {Balaguera-Antolinez}, A. and {Balbinot}, E. and {Balcells}, M. and {Baldi}, M. and {Baldry}, I. and {Balestra}, A. and {Ballardini}, M. and {Ballester}, O. and {Balogh}, M. and {Ba{\~n}ados}, E. and {Barbier}, R. and {Bardelli}, S. and {Baron}, M. and {Barreiro}, T. and {Barrena}, R. and {Barriere}, J.-C. and {Barros}, B.~J. and {Barthelemy}, A. and {Bartolo}, N. and {Basset}, A. and {Battaglia}, P. and {Battisti}, A.~J. and {Baugh}, C.~M. and {Baumont}, L. and {Bazzanini}, L. and {Beaulieu}, J.-P. and {Beckmann}, V. and {Belikov}, A.~N. and {Bel}, J. and {Bellagamba}, F. and {Bella}, M. and {Bellini}, E. and {Benabed}, K. and {Bender}, R. and {Benevento}, G. and {Bennett}, C.~L. and {Benson}, K. and {Bergamini}, P. and {Bermejo-Climent}, J.~R. and {Bernardeau}, F. and {Bertacca}, D. and {Berthe}, M. and {Berthier}, J. and {Bethermin}, M. and {Beutler}, F. and {Bevillon}, C. and {Bhargava}, S. and {Bhatawdekar}, R. and {Bianchi}, D. and {Bisigello}, L. and {Biviano}, A. and {Blake}, R.~P. and {Blanchard}, A. and {Blazek}, J. and {Blot}, L. and {Bosco}, A. and {Bodendorf}, C. and {Boenke}, T. and {B{\"o}hringer}, H. and {Boldrini}, P. and {Bolzonella}, M. and {Bonchi}, A. and {Bonici}, M. and {Bonino}, D. and {Bonino}, L. and {Bonvin}, C. and {Bon}, W. and {Booth}, J.~T. and {Borgani}, S. and {Borlaff}, A.~S. and {Borsato}, E. and {Bose}, B. and {Botticella}, M.~T. and {Boucaud}, A. and {Bouche}, F. and {Boucher}, J.~S. and {Boutigny}, D. and {Bouvard}, T. and {Bouwens}, R. and {Bouy}, H. and {Bowler}, R.~A.~A. and {Bozza}, V. and {Bozzo}, E. and {Branchini}, E. and {Brando}, G. and {Brau-Nogue}, S. and {Brekke}, P. and {Bremer}, M.~N. and {Brescia}, M. and {Breton}, M.-A. and {Brinchmann}, J. and {Brinckmann}, T. and {Brockley-Blatt}, C. and {Brodwin}, M. and {Brouard}, L. and {Brown}, M.~L. and {Bruton}, S. and {Bucko}, J. and {Buddelmeijer}, H. and {Buenadicha}, G. and {Buitrago}, F. and {Burger}, P. and {Burigana}, C. and {Busillo}, V. and {Busonero}, D. and {Cabanac}, R. and {Cabayol-Garcia}, L. and {Cagliari}, M.~S. and {Caillat}, A. and {Caillat}, L. and {Calabrese}, M. and {Calabro}, A. and {Calderone}, G. and {Calura}, F. and {Camacho Quevedo}, B. and {Camera}, S. and {Campos}, L. and {Ca{\~n}as-Herrera}, G. and {Candini}, G.~P. and {Cantiello}, M. and {Capobianco}, V. and {Cappellaro}, E. and {Cappelluti}, N. and {Cappi}, A. and {Caputi}, K.~I. and {Cara}, C. and {Carbone}, C. and {Cardone}, V.~F. and {Carella}, E. and {Carlberg}, R.~G. and {Carle}, M. and {Carminati}, L. and {Caro}, F. and {Carrasco}, J.~M. and {Carretero}, J. and {Carrilho}, P. and {Carron Duque}, J. and {Carry}, B.},
        title = "{Euclid: I. Overview of the Euclid mission}",
      journal = {\aap},
         year = 2025,
        month = may,
       volume = {697},
          eid = {A1},
        pages = {A1},
          doi = {10.1051/0004-6361/202450810},
archivePrefix = {arXiv},
       eprint = {2405.13491},
 primaryClass = {astro-ph.CO},
       adsurl = {https://ui.adsabs.harvard.edu/abs/2025A&A...697A...1E}
}

@ARTICLE{Roman,
       author = {{Spergel}, D. and {Gehrels}, N. and {Baltay}, C. and {Bennett}, D. and {Breckinridge}, J. and {Donahue}, M. and {Dressler}, A. and {Gaudi}, B.~S. and {Greene}, T. and {Guyon}, O. and {Hirata}, C. and {Kalirai}, J. and {Kasdin}, N.~J. and {Macintosh}, B. and {Moos}, W. and {Perlmutter}, S. and {Postman}, M. and {Rauscher}, B. and {Rhodes}, J. and {Wang}, Y. and {Weinberg}, D. and {Benford}, D. and {Hudson}, M. and {Jeong}, W.-S. and {Mellier}, Y. and {Traub}, W. and {Yamada}, T. and {Capak}, P. and {Colbert}, J. and {Masters}, D. and {Penny}, M. and {Savransky}, D. and {Stern}, D. and {Zimmerman}, N. and {Barry}, R. and {Bartusek}, L. and {Carpenter}, K. and {Cheng}, E. and {Content}, D. and {Dekens}, F. and {Demers}, R. and {Grady}, K. and {Jackson}, C. and {Kuan}, G. and {Kruk}, J. and {Melton}, M. and {Nemati}, B. and {Parvin}, B. and {Poberezhskiy}, I. and {Peddie}, C. and {Ruffa}, J. and {Wallace}, J.~K. and {Whipple}, A. and {Wollack}, E. and {Zhao}, F.},
        title = "{Wide-Field InfrarRed Survey Telescope-Astrophysics Focused Telescope Assets WFIRST-AFTA 2015 Report}",
      journal = {arXiv e-prints},
         year = 2015,
        month = mar,
          eid = {arXiv:1503.03757},
        pages = {arXiv:1503.03757},
          doi = {10.48550/arXiv.1503.03757},
archivePrefix = {arXiv},
       eprint = {1503.03757},
 primaryClass = {astro-ph.IM},
       adsurl = {https://ui.adsabs.harvard.edu/abs/2015arXiv150303757S}
}

@ARTICLE{LSST,
       author = {{Ivezi{\'c}}, {\v{Z}}eljko and {Kahn}, Steven M. and {Tyson}, J. Anthony and {Abel}, Bob and {Acosta}, Emily and {Allsman}, Robyn and {Alonso}, David and {AlSayyad}, Yusra and {Anderson}, Scott F. and {Andrew}, John and {Angel}, James Roger P. and {Angeli}, George Z. and {Ansari}, Reza and {Antilogus}, Pierre and {Araujo}, Constanza and {Armstrong}, Robert and {Arndt}, Kirk T. and {Astier}, Pierre and {Aubourg}, {\'E}ric and {Auza}, Nicole and {Axelrod}, Tim S. and {Bard}, Deborah J. and {Barr}, Jeff D. and {Barrau}, Aurelian and {Bartlett}, James G. and {Bauer}, Amanda E. and {Bauman}, Brian J. and {Baumont}, Sylvain and {Bechtol}, Ellen and {Bechtol}, Keith and {Becker}, Andrew C. and {Becla}, Jacek and {Beldica}, Cristina and {Bellavia}, Steve and {Bianco}, Federica B. and {Biswas}, Rahul and {Blanc}, Guillaume and {Blazek}, Jonathan and {Blandford}, Roger D. and {Bloom}, Josh S. and {Bogart}, Joanne and {Bond}, Tim W. and {Booth}, Michael T. and {Borgland}, Anders W. and {Borne}, Kirk and {Bosch}, James F. and {Boutigny}, Dominique and {Brackett}, Craig A. and {Bradshaw}, Andrew and {Brandt}, William Nielsen and {Brown}, Michael E. and {Bullock}, James S. and {Burchat}, Patricia and {Burke}, David L. and {Cagnoli}, Gianpietro and {Calabrese}, Daniel and {Callahan}, Shawn and {Callen}, Alice L. and {Carlin}, Jeffrey L. and {Carlson}, Erin L. and {Chandrasekharan}, Srinivasan and {Charles-Emerson}, Glenaver and {Chesley}, Steve and {Cheu}, Elliott C. and {Chiang}, Hsin-Fang and {Chiang}, James and {Chirino}, Carol and {Chow}, Derek and {Ciardi}, David R. and {Claver}, Charles F. and {Cohen-Tanugi}, Johann and {Cockrum}, Joseph J. and {Coles}, Rebecca and {Connolly}, Andrew J. and {Cook}, Kem H. and {Cooray}, Asantha and {Covey}, Kevin R. and {Cribbs}, Chris and {Cui}, Wei and {Cutri}, Roc and {Daly}, Philip N. and {Daniel}, Scott F. and {Daruich}, Felipe and {Daubard}, Guillaume and {Daues}, Greg and {Dawson}, William and {Delgado}, Francisco and {Dellapenna}, Alfred and {de Peyster}, Robert and {de Val-Borro}, Miguel and {Digel}, Seth W. and {Doherty}, Peter and {Dubois}, Richard and {Dubois-Felsmann}, Gregory P. and {Durech}, Josef and {Economou}, Frossie and {Eifler}, Tim and {Eracleous}, Michael and {Emmons}, Benjamin L. and {Fausti Neto}, Angelo and {Ferguson}, Henry and {Figueroa}, Enrique and {Fisher-Levine}, Merlin and {Focke}, Warren and {Foss}, Michael D. and {Frank}, James and {Freemon}, Michael D. and {Gangler}, Emmanuel and {Gawiser}, Eric and {Geary}, John C. and {Gee}, Perry and {Geha}, Marla and {Gessner}, Charles J.~B. and {Gibson}, Robert R. and {Gilmore}, D. Kirk and {Glanzman}, Thomas and {Glick}, William and {Goldina}, Tatiana and {Goldstein}, Daniel A. and {Goodenow}, Iain and {Graham}, Melissa L. and {Gressler}, William J. and {Gris}, Philippe and {Guy}, Leanne P. and {Guyonnet}, Augustin and {Haller}, Gunther and {Harris}, Ron and {Hascall}, Patrick A. and {Haupt}, Justine and {Hernandez}, Fabio and {Herrmann}, Sven and {Hileman}, Edward and {Hoblitt}, Joshua and {Hodgson}, John A. and {Hogan}, Craig and {Howard}, James D. and {Huang}, Dajun and {Huffer}, Michael E. and {Ingraham}, Patrick and {Innes}, Walter R. and {Jacoby}, Suzanne H. and {Jain}, Bhuvnesh and {Jammes}, Fabrice and {Jee}, M. James and {Jenness}, Tim and {Jernigan}, Garrett and {Jevremovi{\'c}}, Darko and {Johns}, Kenneth and {Johnson}, Anthony S. and {Johnson}, Margaret W.~G. and {Jones}, R. Lynne and {Juramy-Gilles}, Claire and {Juri{\'c}}, Mario and {Kalirai}, Jason S. and {Kallivayalil}, Nitya J. and {Kalmbach}, Bryce and {Kantor}, Jeffrey P. and {Karst}, Pierre and {Kasliwal}, Mansi M. and {Kelly}, Heather and {Kessler}, Richard and {Kinnison}, Veronica and {Kirkby}, David and {Knox}, Lloyd and {Kotov}, Ivan V. and {Krabbendam}, Victor L. and {Krughoff}, K. Simon and {Kub{\'a}nek}, Petr and {Kuczewski}, John and {Kulkarni}, Shri and {Ku}, John and {Kurita}, Nadine R. and {Lage}, Craig S. and {Lambert}, Ron and {Lange}, Travis and {Langton}, J. Brian and {Le Guillou}, Laurent and {Levine}, Deborah and {Liang}, Ming and {Lim}, Kian-Tat and {Lintott}, Chris J. and {Long}, Kevin E. and {Lopez}, Margaux and {Lotz}, Paul J. and {Lupton}, Robert H. and {Lust}, Nate B. and {MacArthur}, Lauren A. and {Mahabal}, Ashish and {Mandelbaum}, Rachel and {Markiewicz}, Thomas W. and {Marsh}, Darren S. and {Marshall}, Philip J. and {Marshall}, Stuart and {May}, Morgan and {McKercher}, Robert and {McQueen}, Michelle and {Meyers}, Joshua and {Migliore}, Myriam and {Miller}, Michelle and {Mills}, David J.},
        title = "{LSST: From Science Drivers to Reference Design and Anticipated Data Products}",
      journal = {\apj},
         year = 2019,
        month = mar,
       volume = {873},
       number = {2},
          eid = {111},
        pages = {111},
          doi = {10.3847/1538-4357/ab042c},
archivePrefix = {arXiv},
       eprint = {0805.2366},
 primaryClass = {astro-ph},
       adsurl = {https://ui.adsabs.harvard.edu/abs/2019ApJ...873..111I}
}

@ARTICLE{Wechsler&Tinker,
       author = {{Wechsler}, Risa H. and {Tinker}, Jeremy L.},
        title = "{The Connection Between Galaxies and Their Dark Matter Halos}",
      journal = {\araa},
         year = 2018,
        month = sep,
       volume = {56},
        pages = {435-487},
          doi = {10.1146/annurev-astro-081817-051756},
archivePrefix = {arXiv},
       eprint = {1804.03097},
 primaryClass = {astro-ph.GA},
       adsurl = {https://ui.adsabs.harvard.edu/abs/2018ARA&A..56..435W}
}

@ARTICLE{Illustris,
       author = {{Vogelsberger}, Mark and {Genel}, Shy and {Springel}, Volker and {Torrey}, Paul and {Sijacki}, Debora and {Xu}, Dandan and {Snyder}, Greg and {Nelson}, Dylan and {Hernquist}, Lars},
        title = "{Introducing the Illustris Project: simulating the coevolution of dark and visible matter in the Universe}",
      journal = {\mnras},
         year = 2014,
        month = oct,
       volume = {444},
       number = {2},
        pages = {1518-1547},
          doi = {10.1093/mnras/stu1536},
archivePrefix = {arXiv},
       eprint = {1405.2921},
 primaryClass = {astro-ph.CO},
       adsurl = {https://ui.adsabs.harvard.edu/abs/2014MNRAS.444.1518V}
}

@ARTICLE{EAGLE,
       author = {{Schaye}, Joop and {Crain}, Robert A. and {Bower}, Richard G. and {Furlong}, Michelle and {Schaller}, Matthieu and {Theuns}, Tom and {Dalla Vecchia}, Claudio and {Frenk}, Carlos S. and {McCarthy}, I.~G. and {Helly}, John C. and {Jenkins}, Adrian and {Rosas-Guevara}, Y.~M. and {White}, Simon D.~M. and {Baes}, Maarten and {Booth}, C.~M. and {Camps}, Peter and {Navarro}, Julio F. and {Qu}, Yan and {Rahmati}, Alireza and {Sawala}, Till and {Thomas}, Peter A. and {Trayford}, James},
        title = "{The EAGLE project: simulating the evolution and assembly of galaxies and their environments}",
      journal = {\mnras},
         year = 2015,
        month = jan,
       volume = {446},
       number = {1},
        pages = {521-554},
          doi = {10.1093/mnras/stu2058},
archivePrefix = {arXiv},
       eprint = {1407.7040},
 primaryClass = {astro-ph.GA},
       adsurl = {https://ui.adsabs.harvard.edu/abs/2015MNRAS.446..521S}
}

@ARTICLE{Flamingo,
       author = {{Schaye}, Joop and {Kugel}, Roi and {Schaller}, Matthieu and {Helly}, John C. and {Braspenning}, Joey and {Elbers}, Willem and {McCarthy}, Ian G. and {van Daalen}, Marcel P. and {Vandenbroucke}, Bert and {Frenk}, Carlos S. and {Kwan}, Juliana and {Salcido}, Jaime and {Bah{\'e}}, Yannick M. and {Borrow}, Josh and {Chaikin}, Evgenii and {Hahn}, Oliver and {Hu{\v{s}}ko}, Filip and {Jenkins}, Adrian and {Lacey}, Cedric G. and {Nobels}, Folkert S.~J.},
        title = "{The FLAMINGO project: cosmological hydrodynamical simulations for large-scale structure and galaxy cluster surveys}",
      journal = {\mnras},
         year = 2023,
        month = dec,
       volume = {526},
       number = {4},
        pages = {4978-5020},
          doi = {10.1093/mnras/stad2419},
archivePrefix = {arXiv},
       eprint = {2306.04024},
 primaryClass = {astro-ph.CO},
       adsurl = {https://ui.adsabs.harvard.edu/abs/2023MNRAS.526.4978S}
}

@ARTICLE{Somerville2008,
       author = {{Somerville}, Rachel S. and {Hopkins}, Philip F. and {Cox}, Thomas J. and {Robertson}, Brant E. and {Hernquist}, Lars},
        title = "{A semi-analytic model for the co-evolution of galaxies, black holes and active galactic nuclei}",
      journal = {\mnras},
         year = 2008,
        month = dec,
       volume = {391},
       number = {2},
        pages = {481-506},
          doi = {10.1111/j.1365-2966.2008.13805.x},
archivePrefix = {arXiv},
       eprint = {0808.1227},
 primaryClass = {astro-ph},
       adsurl = {https://ui.adsabs.harvard.edu/abs/2008MNRAS.391..481S}
}

@ARTICLE{eifler_krause_2017_cosmolike,
       author = {{Krause}, Elisabeth and {Eifler}, Tim},
        title = "{cosmolike - cosmological likelihood analyses for photometric galaxy surveys}",
      journal = {\mnras},
         year = 2017,
        month = sep,
       volume = {470},
       number = {2},
        pages = {2100-2112},
          doi = {10.1093/mnras/stx1261},
archivePrefix = {arXiv},
       eprint = {1601.05779},
 primaryClass = {astro-ph.CO},
       adsurl = {https://ui.adsabs.harvard.edu/abs/2017MNRAS.470.2100K}
}

@ARTICLE{Somerville2015,
       author = {{Somerville}, Rachel S. and {Popping}, Gerg{\"o} and {Trager}, Scott C.},
        title = "{Star formation in semi-analytic galaxy formation models with multiphase gas}",
      journal = {\mnras},
         year = 2015,
        month = nov,
       volume = {453},
       number = {4},
        pages = {4337-4367},
          doi = {10.1093/mnras/stv1877},
archivePrefix = {arXiv},
       eprint = {1503.00755},
 primaryClass = {astro-ph.GA},
       adsurl = {https://ui.adsabs.harvard.edu/abs/2015MNRAS.453.4337S}
}

@ARTICLE{Henrique2015,
       author = {{Henriques}, Bruno M.~B. and {White}, Simon D.~M. and {Thomas}, Peter A. and {Angulo}, Raul and {Guo}, Qi and {Lemson}, Gerard and {Springel}, Volker and {Overzier}, Roderik},
        title = "{Galaxy formation in the Planck cosmology - I. Matching the observed evolution of star formation rates, colours and stellar masses}",
      journal = {\mnras},
         year = 2015,
        month = aug,
       volume = {451},
       number = {3},
        pages = {2663-2680},
          doi = {10.1093/mnras/stv705},
archivePrefix = {arXiv},
       eprint = {1410.0365},
 primaryClass = {astro-ph.GA},
       adsurl = {https://ui.adsabs.harvard.edu/abs/2015MNRAS.451.2663H}
}

@ARTICLE{Henrique2020,
       author = {{Henriques}, Bruno M.~B. and {Yates}, Robert M. and {Fu}, Jian and {Guo}, Qi and {Kauffmann}, Guinevere and {Srisawat}, Chaichalit and {Thomas}, Peter A. and {White}, Simon D.~M.},
        title = "{L-GALAXIES 2020: Spatially resolved cold gas phases, star formation, and chemical enrichment in galactic discs}",
      journal = {\mnras},
         year = 2020,
        month = feb,
       volume = {491},
       number = {4},
        pages = {5795-5814},
          doi = {10.1093/mnras/stz3233},
archivePrefix = {arXiv},
       eprint = {2003.05944},
 primaryClass = {astro-ph.GA},
       adsurl = {https://ui.adsabs.harvard.edu/abs/2020MNRAS.491.5795H}
}

@ARTICLE{Lagos2024,
       author = {{Lagos}, Claudia del P. and {Bravo}, Mat{\'\i}as and {Tobar}, Rodrigo and {Obreschkow}, Danail and {Power}, Chris and {Robotham}, Aaron S.~G. and {Proctor}, Katy L. and {Hansen}, Samuel and {Chandro-G{\'o}mez}, {\'A}ngel and {Carrivick}, Julian},
        title = "{Quenching massive galaxies across cosmic time with the semi-analytic model SHARK V2.0}",
      journal = {\mnras},
         year = 2024,
        month = jul,
       volume = {531},
       number = {3},
        pages = {3551-3578},
          doi = {10.1093/mnras/stae1024},
archivePrefix = {arXiv},
       eprint = {2309.02310},
 primaryClass = {astro-ph.GA},
       adsurl = {https://ui.adsabs.harvard.edu/abs/2024MNRAS.531.3551L}
}

@ARTICLE{EMERGE,
       author = {{Moster}, Benjamin P. and {Naab}, Thorsten and {White}, Simon D.~M.},
        title = "{EMERGE - an empirical model for the formation of galaxies since z {\ensuremath{\sim}} 10}",
      journal = {\mnras},
         year = 2018,
        month = jun,
       volume = {477},
       number = {2},
        pages = {1822-1852},
          doi = {10.1093/mnras/sty655},
archivePrefix = {arXiv},
       eprint = {1705.05373},
 primaryClass = {astro-ph.GA},
       adsurl = {https://ui.adsabs.harvard.edu/abs/2018MNRAS.477.1822M}
}

@ARTICLE{Diffstar,
      author = {{Alarcon}, Alex and {Hearin}, Andrew P. and {Becker}, Matthew R. and {Chaves-Montero}, Jon{\'a}s},
        title = "{Diffstar: a fully parametric physical model for galaxy assembly history}",
      journal = {MNRAS},
        year = 2023,
        month = jan,
      volume = {518},
      number = {1},
        pages = {562-584},
          doi = {10.1093/mnras/stac3118},
archivePrefix = {arXiv},
      eprint = {2205.04273},
primaryClass = {astro-ph.GA},
      adsurl = {https://ui.adsabs.harvard.edu/abs/2023MNRAS.518..562A}
}

@ARTICLE{Kravtsov2004,
       author = {{Kravtsov}, Andrey V. and {Berlind}, Andreas A. and {Wechsler}, Risa H. and {Klypin}, Anatoly A. and {Gottl{\"o}ber}, Stefan and {Allgood}, Brandon and {Primack}, Joel R.},
        title = "{The Dark Side of the Halo Occupation Distribution}",
      journal = {\apj},
         year = 2004,
        month = jul,
       volume = {609},
       number = {1},
        pages = {35-49},
          doi = {10.1086/420959},
archivePrefix = {arXiv},
       eprint = {astro-ph/0308519},
 primaryClass = {astro-ph},
       adsurl = {https://ui.adsabs.harvard.edu/abs/2004ApJ...609...35K}
}

@ARTICLE{Vale2006,
       author = {{Vale}, A. and {Ostriker}, J.~P.},
        title = "{The non-parametric model for linking galaxy luminosity with halo/subhalo mass}",
      journal = {\mnras},
         year = 2006,
        month = sep,
       volume = {371},
       number = {3},
        pages = {1173-1187},
          doi = {10.1111/j.1365-2966.2006.10605.x},
archivePrefix = {arXiv},
       eprint = {astro-ph/0511816},
 primaryClass = {astro-ph},
       adsurl = {https://ui.adsabs.harvard.edu/abs/2006MNRAS.371.1173V}
}

@ARTICLE{Conroy2006,
       author = {{Conroy}, Charlie and {Wechsler}, Risa H. and {Kravtsov}, Andrey V.},
        title = "{Modeling Luminosity-dependent Galaxy Clustering through Cosmic Time}",
      journal = {\apj},
         year = 2006,
        month = aug,
       volume = {647},
       number = {1},
        pages = {201-214},
          doi = {10.1086/503602},
archivePrefix = {arXiv},
       eprint = {astro-ph/0512234},
 primaryClass = {astro-ph},
       adsurl = {https://ui.adsabs.harvard.edu/abs/2006ApJ...647..201C}
}

@ARTICLE{Peacock2000,
       author = {{Peacock}, J.~A. and {Smith}, R.~E.},
        title = "{Halo occupation numbers and galaxy bias}",
      journal = {\mnras},
         year = 2000,
        month = nov,
       volume = {318},
       number = {4},
        pages = {1144-1156},
          doi = {10.1046/j.1365-8711.2000.03779.x},
archivePrefix = {arXiv},
       eprint = {astro-ph/0005010},
 primaryClass = {astro-ph},
       adsurl = {https://ui.adsabs.harvard.edu/abs/2000MNRAS.318.1144P}
}

@ARTICLE{Seljak2000,
       author = {{Seljak}, Uro{\v{s}}},
        title = "{Analytic model for galaxy and dark matter clustering}",
      journal = {\mnras},
         year = 2000,
        month = oct,
       volume = {318},
       number = {1},
        pages = {203-213},
          doi = {10.1046/j.1365-8711.2000.03715.x},
archivePrefix = {arXiv},
       eprint = {astro-ph/0001493},
 primaryClass = {astro-ph},
       adsurl = {https://ui.adsabs.harvard.edu/abs/2000MNRAS.318..203S}
}

@ARTICLE{Zheng2005,
       author = {{Zheng}, Zheng and {Berlind}, Andreas A. and {Weinberg}, David H. and {Benson}, Andrew J. and {Baugh}, Carlton M. and {Cole}, Shaun and {Dav{\'e}}, Romeel and {Frenk}, Carlos S. and {Katz}, Neal and {Lacey}, Cedric G.},
        title = "{Theoretical Models of the Halo Occupation Distribution: Separating Central and Satellite Galaxies}",
      journal = {\apj},
         year = 2005,
        month = nov,
       volume = {633},
       number = {2},
        pages = {791-809},
          doi = {10.1086/466510},
archivePrefix = {arXiv},
       eprint = {astro-ph/0408564},
 primaryClass = {astro-ph},
       adsurl = {https://ui.adsabs.harvard.edu/abs/2005ApJ...633..791Z}
}

@ARTICLE{Xu2021,
       author = {{Xu}, Xiaoju and {Zehavi}, Idit and {Contreras}, Sergio},
        title = "{Dissecting and modelling galaxy assembly bias}",
      journal = {\mnras},
         year = 2021,
        month = apr,
       volume = {502},
       number = {3},
        pages = {3242-3263},
          doi = {10.1093/mnras/stab100},
archivePrefix = {arXiv},
       eprint = {2007.05545},
 primaryClass = {astro-ph.GA},
       adsurl = {https://ui.adsabs.harvard.edu/abs/2021MNRAS.502.3242X}
}

@ARTICLE{Rockstar,
       author = {{Behroozi}, Peter S. and {Wechsler}, Risa H. and {Wu}, Hao-Yi},
        title = "{The ROCKSTAR Phase-space Temporal Halo Finder and the Velocity Offsets of Cluster Cores}",
      journal = {\apj},
         year = 2013,
        month = jan,
       volume = {762},
       number = {2},
          eid = {109},
        pages = {109},
          doi = {10.1088/0004-637X/762/2/109},
archivePrefix = {arXiv},
       eprint = {1110.4372},
 primaryClass = {astro-ph.CO},
       adsurl = {https://ui.adsabs.harvard.edu/abs/2013ApJ...762..109B}
}

@ARTICLE{ctree,
       author = {{Behroozi}, Peter S. and {Wechsler}, Risa H. and {Wu}, Hao-Yi and {Busha}, Michael T. and {Klypin}, Anatoly A. and {Primack}, Joel R.},
        title = "{Gravitationally Consistent Halo Catalogs and Merger Trees for Precision Cosmology}",
      journal = {\apj},
         year = 2013,
        month = jan,
       volume = {763},
       number = {1},
          eid = {18},
        pages = {18},
          doi = {10.1088/0004-637X/763/1/18},
archivePrefix = {arXiv},
       eprint = {1110.4370},
 primaryClass = {astro-ph.CO},
       adsurl = {https://ui.adsabs.harvard.edu/abs/2013ApJ...763...18B}
}

@ARTICLE{Lehmann2017,
       author = {{Lehmann}, Benjamin V. and {Mao}, Yao-Yuan and {Becker}, Matthew R. and {Skillman}, Samuel W. and {Wechsler}, Risa H.},
        title = "{The Concentration Dependence of the Galaxy-Halo Connection: Modeling Assembly Bias with Abundance Matching}",
      journal = {\apj},
         year = 2017,
        month = jan,
       volume = {834},
       number = {1},
          eid = {37},
        pages = {37},
          doi = {10.3847/1538-4357/834/1/37},
archivePrefix = {arXiv},
       eprint = {1510.05651},
 primaryClass = {astro-ph.CO},
       adsurl = {https://ui.adsabs.harvard.edu/abs/2017ApJ...834...37L}
}

@ARTICLE{Zheng07,
       author = {{Zheng}, Zheng and {Coil}, Alison L. and {Zehavi}, Idit},
        title = "{Galaxy Evolution from Halo Occupation Distribution Modeling of DEEP2 and SDSS Galaxy Clustering}",
      journal = {\apj},
         year = 2007,
        month = oct,
       volume = {667},
       number = {2},
        pages = {760-779},
          doi = {10.1086/521074},
archivePrefix = {arXiv},
       eprint = {astro-ph/0703457},
 primaryClass = {astro-ph},
       adsurl = {https://ui.adsabs.harvard.edu/abs/2007ApJ...667..760Z}
}

@ARTICLE{Hearin2016,
       author = {{Hearin}, Andrew P. and {Zentner}, Andrew R. and {van den Bosch}, Frank C. and {Campbell}, Duncan and {Tollerud}, Erik},
        title = "{Introducing decorated HODs: modelling assembly bias in the galaxy-halo connection}",
      journal = {\mnras},
         year = 2016,
        month = aug,
       volume = {460},
       number = {3},
        pages = {2552-2570},
          doi = {10.1093/mnras/stw840},
archivePrefix = {arXiv},
       eprint = {1512.03050},
 primaryClass = {astro-ph.CO},
       adsurl = {https://ui.adsabs.harvard.edu/abs/2016MNRAS.460.2552H}
}

@ARTICLE{CEM,
       author = {{Lange}, Johannes U. and {van den Bosch}, Frank C. and {Zentner}, Andrew R. and {Wang}, Kuan and {Hearin}, Andrew P. and {Guo}, Hong},
        title = "{Cosmological Evidence Modelling: a new simulation-based approach to constrain cosmology on non-linear scales}",
      journal = {\mnras},
         year = 2019,
        month = dec,
       volume = {490},
       number = {2},
        pages = {1870-1878},
          doi = {10.1093/mnras/stz2664},
archivePrefix = {arXiv},
       eprint = {1909.03107},
 primaryClass = {astro-ph.CO},
       adsurl = {https://ui.adsabs.harvard.edu/abs/2019MNRAS.490.1870L}
}

@ARTICLE{AbacusHOD,
       author = {{Yuan}, Sihan and {Garrison}, Lehman H. and {Hadzhiyska}, Boryana and {Bose}, Sownak and {Eisenstein}, Daniel J.},
        title = "{ABACUSHOD: a highly efficient extended multitracer HOD framework and its application to BOSS and eBOSS data}",
      journal = {\mnras},
         year = 2022,
        month = mar,
       volume = {510},
       number = {3},
        pages = {3301-3320},
          doi = {10.1093/mnras/stab3355},
archivePrefix = {arXiv},
       eprint = {2110.11412},
 primaryClass = {astro-ph.CO},
       adsurl = {https://ui.adsabs.harvard.edu/abs/2022MNRAS.510.3301Y}
}

@ARTICLE{halotools,
       author = {{Hearin}, Andrew P. and {Campbell}, Duncan and {Tollerud}, Erik and {Behroozi}, Peter and {Diemer}, Benedikt and {Goldbaum}, Nathan J. and {Jennings}, Elise and {Leauthaud}, Alexie and {Mao}, Yao-Yuan and {More}, Surhud and {Parejko}, John and {Sinha}, Manodeep and {Sip{\"o}cz}, Brigitta and {Zentner}, Andrew},
        title = "{Forward Modeling of Large-scale Structure: An Open-source Approach with Halotools}",
      journal = {\aj},
         year = 2017,
        month = nov,
       volume = {154},
       number = {5},
          eid = {190},
        pages = {190},
          doi = {10.3847/1538-3881/aa859f},
archivePrefix = {arXiv},
       eprint = {1606.04106},
 primaryClass = {astro-ph.IM},
       adsurl = {https://ui.adsabs.harvard.edu/abs/2017AJ....154..190H}
}

@ARTICLE{Wechsler2002,
       author = {{Wechsler}, Risa H. and {Bullock}, James S. and {Primack}, Joel R. and {Kravtsov}, Andrey V. and {Dekel}, Avishai},
        title = "{Concentrations of Dark Halos from Their Assembly Histories}",
      journal = {\apj},
         year = 2002,
        month = mar,
       volume = {568},
       number = {1},
        pages = {52-70},
          doi = {10.1086/338765},
archivePrefix = {arXiv},
       eprint = {astro-ph/0108151},
 primaryClass = {astro-ph},
       adsurl = {https://ui.adsabs.harvard.edu/abs/2002ApJ...568...52W}
}

@ARTICLE{Wechsler2006,
       author = {{Wechsler}, Risa H. and {Zentner}, Andrew R. and {Bullock}, James S. and {Kravtsov}, Andrey V. and {Allgood}, Brandon},
        title = "{The Dependence of Halo Clustering on Halo Formation History, Concentration, and Occupation}",
      journal = {\apj},
         year = 2006,
        month = nov,
       volume = {652},
       number = {1},
        pages = {71-84},
          doi = {10.1086/507120},
archivePrefix = {arXiv},
       eprint = {astro-ph/0512416},
 primaryClass = {astro-ph},
       adsurl = {https://ui.adsabs.harvard.edu/abs/2006ApJ...652...71W}
}

@ARTICLE{Hadzhiyska2020,
       author = {{Hadzhiyska}, Boryana and {Bose}, Sownak and {Eisenstein}, Daniel and {Hernquist}, Lars and {Spergel}, David N.},
        title = "{Limitations to the `basic' HOD model and beyond}",
      journal = {\mnras},
         year = 2020,
        month = apr,
       volume = {493},
       number = {4},
        pages = {5506-5519},
          doi = {10.1093/mnras/staa623},
archivePrefix = {arXiv},
       eprint = {1911.02610},
 primaryClass = {astro-ph.CO},
       adsurl = {https://ui.adsabs.harvard.edu/abs/2020MNRAS.493.5506H}
}

@ARTICLE{Corrfunc,
       author = {{Sinha}, Manodeep and {Garrison}, Lehman H.},
        title = "{CORRFUNC - a suite of blazing fast correlation functions on the CPU}",
      journal = {\mnras},
         year = 2020,
        month = jan,
       volume = {491},
       number = {2},
        pages = {3022-3041},
          doi = {10.1093/mnras/stz3157},
archivePrefix = {arXiv},
       eprint = {1911.03545},
 primaryClass = {astro-ph.CO},
       adsurl = {https://ui.adsabs.harvard.edu/abs/2020MNRAS.491.3022S}
}

@ARTICLE{Nautilus,
       author = {{Lange}, Johannes U.},
        title = "{NAUTILUS: boosting Bayesian importance nested sampling with deep learning}",
      journal = {\mnras},
         year = 2023,
        month = oct,
       volume = {525},
       number = {2},
        pages = {3181-3194},
          doi = {10.1093/mnras/stad2441},
archivePrefix = {arXiv},
       eprint = {2306.16923},
 primaryClass = {astro-ph.IM},
       adsurl = {https://ui.adsabs.harvard.edu/abs/2023MNRAS.525.3181L}
}

@ARTICLE{Splashback,
       author = {{Mansfield}, Philip and {Kravtsov}, Andrey V.},
        title = "{The three causes of low-mass assembly bias}",
      journal = {\mnras},
         year = 2020,
        month = apr,
       volume = {493},
       number = {4},
        pages = {4763-4782},
          doi = {10.1093/mnras/staa430},
archivePrefix = {arXiv},
       eprint = {1902.00030},
 primaryClass = {astro-ph.CO},
       adsurl = {https://ui.adsabs.harvard.edu/abs/2020MNRAS.493.4763M}
}

@ARTICLE{UniverseMachine,
       author = {{Behroozi}, Peter and {Wechsler}, Risa H. and {Hearin}, Andrew P. and {Conroy}, Charlie},
        title = "{UNIVERSEMACHINE: The correlation between galaxy growth and dark matter halo assembly from z = 0-10}",
      journal = {\mnras},
         year = 2019,
        month = sep,
       volume = {488},
       number = {3},
        pages = {3143-3194},
          doi = {10.1093/mnras/stz1182},
archivePrefix = {arXiv},
       eprint = {1806.07893},
 primaryClass = {astro-ph.GA},
       adsurl = {https://ui.adsabs.harvard.edu/abs/2019MNRAS.488.3143B}
}

@ARTICLE{Planck2018,
       author = {{Planck Collaboration} and {Aghanim}, N. and {Akrami}, Y. and {Ashdown}, M. and {Aumont}, J. and {Baccigalupi}, C. and {Ballardini}, M. and {Banday}, A.~J. and {Barreiro}, R.~B. and {Bartolo}, N. and {Basak}, S. and {Battye}, R. and {Benabed}, K. and {Bernard}, J. -P. and {Bersanelli}, M. and {Bielewicz}, P. and {Bock}, J.~J. and {Bond}, J.~R. and {Borrill}, J. and {Bouchet}, F.~R. and {Boulanger}, F. and {Bucher}, M. and {Burigana}, C. and {Butler}, R.~C. and {Calabrese}, E. and {Cardoso}, J. -F. and {Carron}, J. and {Challinor}, A. and {Chiang}, H.~C. and {Chluba}, J. and {Colombo}, L.~P.~L. and {Combet}, C. and {Contreras}, D. and {Crill}, B.~P. and {Cuttaia}, F. and {de Bernardis}, P. and {de Zotti}, G. and {Delabrouille}, J. and {Delouis}, J. -M. and {Di Valentino}, E. and {Diego}, J.~M. and {Dor{\'e}}, O. and {Douspis}, M. and {Ducout}, A. and {Dupac}, X. and {Dusini}, S. and {Efstathiou}, G. and {Elsner}, F. and {En{\ss}lin}, T.~A. and {Eriksen}, H.~K. and {Fantaye}, Y. and {Farhang}, M. and {Fergusson}, J. and {Fernandez-Cobos}, R. and {Finelli}, F. and {Forastieri}, F. and {Frailis}, M. and {Fraisse}, A.~A. and {Franceschi}, E. and {Frolov}, A. and {Galeotta}, S. and {Galli}, S. and {Ganga}, K. and {G{\'e}nova-Santos}, R.~T. and {Gerbino}, M. and {Ghosh}, T. and {Gonz{\'a}lez-Nuevo}, J. and {G{\'o}rski}, K.~M. and {Gratton}, S. and {Gruppuso}, A. and {Gudmundsson}, J.~E. and {Hamann}, J. and {Handley}, W. and {Hansen}, F.~K. and {Herranz}, D. and {Hildebrandt}, S.~R. and {Hivon}, E. and {Huang}, Z. and {Jaffe}, A.~H. and {Jones}, W.~C. and {Karakci}, A. and {Keih{\"a}nen}, E. and {Keskitalo}, R. and {Kiiveri}, K. and {Kim}, J. and {Kisner}, T.~S. and {Knox}, L. and {Krachmalnicoff}, N. and {Kunz}, M. and {Kurki-Suonio}, H. and {Lagache}, G. and {Lamarre}, J. -M. and {Lasenby}, A. and {Lattanzi}, M. and {Lawrence}, C.~R. and {Le Jeune}, M. and {Lemos}, P. and {Lesgourgues}, J. and {Levrier}, F. and {Lewis}, A. and {Liguori}, M. and {Lilje}, P.~B. and {Lilley}, M. and {Lindholm}, V. and {L{\'o}pez-Caniego}, M. and {Lubin}, P.~M. and {Ma}, Y. -Z. and {Mac{\'\i}as-P{\'e}rez}, J.~F. and {Maggio}, G. and {Maino}, D. and {Mandolesi}, N. and {Mangilli}, A. and {Marcos-Caballero}, A. and {Maris}, M. and {Martin}, P.~G. and {Martinelli}, M. and {Mart{\'\i}nez-Gonz{\'a}lez}, E. and {Matarrese}, S. and {Mauri}, N. and {McEwen}, J.~D. and {Meinhold}, P.~R. and {Melchiorri}, A. and {Mennella}, A. and {Migliaccio}, M. and {Millea}, M. and {Mitra}, S. and {Miville-Desch{\^e}nes}, M. -A. and {Molinari}, D. and {Montier}, L. and {Morgante}, G. and {Moss}, A. and {Natoli}, P. and {N{\o}rgaard-Nielsen}, H.~U. and {Pagano}, L. and {Paoletti}, D. and {Partridge}, B. and {Patanchon}, G. and {Peiris}, H.~V. and {Perrotta}, F. and {Pettorino}, V. and {Piacentini}, F. and {Polastri}, L. and {Polenta}, G. and {Puget}, J. -L. and {Rachen}, J.~P. and {Reinecke}, M. and {Remazeilles}, M. and {Renzi}, A. and {Rocha}, G. and {Rosset}, C. and {Roudier}, G. and {Rubi{\~n}o-Mart{\'\i}n}, J.~A. and {Ruiz-Granados}, B. and {Salvati}, L. and {Sandri}, M. and {Savelainen}, M. and {Scott}, D. and {Shellard}, E.~P.~S. and {Sirignano}, C. and {Sirri}, G. and {Spencer}, L.~D. and {Sunyaev}, R. and {Suur-Uski}, A. -S. and {Tauber}, J.~A. and {Tavagnacco}, D. and {Tenti}, M. and {Toffolatti}, L. and {Tomasi}, M. and {Trombetti}, T. and {Valenziano}, L. and {Valiviita}, J. and {Van Tent}, B. and {Vibert}, L. and {Vielva}, P. and {Villa}, F. and {Vittorio}, N. and {Wandelt}, B.~D. and {Wehus}, I.~K. and {White}, M. and {White}, S.~D.~M. and {Zacchei}, A. and {Zonca}, A.},
        title = "{Planck 2018 results. VI. Cosmological parameters}",
      journal = {\aap},
         year = 2020,
        month = sep,
       volume = {641},
          eid = {A6},
        pages = {A6},
          doi = {10.1051/0004-6361/201833910},
archivePrefix = {arXiv},
       eprint = {1807.06209},
 primaryClass = {astro-ph.CO},
       adsurl = {https://ui.adsabs.harvard.edu/abs/2020A&A...641A...6P}
}

@ARTICLE{multiplet,
       author = {{Wang}, Hanyue and {Eisenstein}, Daniel J. and {Aguilar}, Jessica Nicole and {Ahlen}, Steven and {Bianchi}, Davide and {Brooks}, David and {Claybaugh}, Todd and {de la Macorra}, Axel and {Dey}, Arjun and {Dey}, Biprateep and {Doel}, Peter and {Ferraro}, Simone and {Font-Ribera}, Andreu and {Forero-Romero}, Jaime E. and {Gazta{\~n}aga}, Enrique and {Gutierrez}, Gaston and {Honscheid}, Klaus and {Ishak}, Mustapha and {Joyce}, Richard and {Juneau}, Stephanie and {Kirkby}, David and {Kisner}, Theodore and {Kremin}, Anthony and {Lahav}, Ofer and {Lamman}, Claire and {Landriau}, Martin and {Manera}, Marc and {Meisner}, Aaron and {Miquel}, Ramon and {Mueller}, Eva-Maria and {Nadathur}, Seshadri and {Niz}, Gustavo and {Palanque-Delabrouille}, Nathalie and {Percival}, Will J. and {Prada}, Francisco and {P{\'e}rez-R{\`a}fols}, Ignasi and {Ross}, Ashley J. and {Rossi}, Graziano and {Sanchez}, Eusebio and {Schlegel}, David and {Schubnell}, Michael and {Silber}, Joseph Harry and {Sprayberry}, David and {Tarl{\'e}}, Gregory and {Weaver}, Benjamin Alan and {Zhou}, Rongpu and {Zou}, Hu},
        title = "{Galaxy-multiplet clustering from DESI DR2}",
      journal = {\mnras},
         year = 2026,
        month = feb,
       volume = {545},
       number = {4},
          eid = {staf2069},
        pages = {staf2069},
          doi = {10.1093/mnras/staf2069},
       adsurl = {https://ui.adsabs.harvard.edu/abs/2026MNRAS.545S2069W}
}

@ARTICLE{SHAMe,
       author = {{Contreras}, S. and {Angulo}, R.~E. and {Zennaro}, M.},
        title = "{A flexible subhalo abundance matching model for galaxy clustering in redshift space}",
      journal = {\mnras},
         year = 2021,
        month = nov,
       volume = {508},
       number = {1},
        pages = {175-189},
          doi = {10.1093/mnras/stab2560},
archivePrefix = {arXiv},
       eprint = {2012.06596},
 primaryClass = {astro-ph.CO},
       adsurl = {https://ui.adsabs.harvard.edu/abs/2021MNRAS.508..175C}
}

@ARTICLE{Guo&White2014,
       author = {{Guo}, Qi and {White}, Simon},
        title = "{Numerical resolution limits on subhalo abundance matching}",
      journal = {\mnras},
         year = 2014,
        month = feb,
       volume = {437},
       number = {4},
        pages = {3228-3235},
          doi = {10.1093/mnras/stt2116},
archivePrefix = {arXiv},
       eprint = {1303.3586},
 primaryClass = {astro-ph.CO},
       adsurl = {https://ui.adsabs.harvard.edu/abs/2014MNRAS.437.3228G}
}

@ARTICLE{DSPS,
      author = {{Hearin}, Andrew P. and {Chaves-Montero}, Jon{\'a}s and {Alarcon}, Alex and {Becker}, Matthew R. and {Benson}, Andrew},
        title = "{DSPS: Differentiable stellar population synthesis}",
      journal = {MNRAS},
        year = 2023,
        month = may,
      volume = {521},
      number = {2},
        pages = {1741-1756},
          doi = {10.1093/mnras/stad456},
archivePrefix = {arXiv},
      eprint = {2112.06830},
primaryClass = {astro-ph.GA},
      adsurl = {https://ui.adsabs.harvard.edu/abs/2023MNRAS.521.1741H}
}

@ARTICLE{Diffstarpop,
       author = {{Alarcon}, Alex and {Hearin}, Andrew P. and {Becker}, Matthew R. and {Beltz-Mohrmann}, Gillian and {Benson}, Andrew and {Weerasooriya}, Sachi},
        title = "{DiffstarPop: A generative physical model of galaxy star formation history}",
      journal = {The Open Journal of Astrophysics},
         year = 2026,
        month = may,
       volume = {9},
        pages = {61920},
          doi = {10.33232/001c.161920},
archivePrefix = {arXiv},
       eprint = {2510.27604},
 primaryClass = {astro-ph.GA},
       adsurl = {https://ui.adsabs.harvard.edu/abs/2026OJAp....961920A}
}

@ARTICLE{Mohammad&Percival,
       author = {{Mohammad}, Faizan G. and {Percival}, Will J.},
        title = "{Creating jackknife and bootstrap estimates of the covariance matrix for the two-point correlation function}",
      journal = {\mnras},
         year = 2022,
        month = jul,
       volume = {514},
       number = {1},
        pages = {1289-1301},
          doi = {10.1093/mnras/stac1458},
archivePrefix = {arXiv},
       eprint = {2109.07071},
 primaryClass = {astro-ph.CO},
       adsurl = {https://ui.adsabs.harvard.edu/abs/2022MNRAS.514.1289M}
}

@ARTICLE{Contreras2021,
       author = {{Contreras}, Sergio and {Angulo}, Raul E. and {Springel}, Volker and {White}, Simon D.~M. and {Hadzhiyska}, Boryana and {Hernquist}, Lars and {Pakmor}, R{\"u}diger and {Kannan}, Rahul and {Hern{\'a}ndez-Aguayo}, C{\'e}sar and {Barrera}, Monica and {Ferlito}, Fulvio and {Delgado}, Ana Maria and {Bose}, Sownak and {Frenk}, Carlos},
        title = "{The MillenniumTNG Project: inferring cosmology from galaxy clustering with accelerated N-body scaling and subhalo abundance matching}",
      journal = {\mnras},
         year = 2023,
        month = sep,
       volume = {524},
       number = {2},
        pages = {2489-2506},
          doi = {10.1093/mnras/stac3699},
archivePrefix = {arXiv},
       eprint = {2210.10075},
 primaryClass = {astro-ph.GA},
       adsurl = {https://ui.adsabs.harvard.edu/abs/2023MNRAS.524.2489C}
}

@ARTICLE{Contreras2026,
       author = {{Mahony}, Constance and {Contreras}, Sergio and {Angulo}, Raul E. and {Alonso}, David and {Georgiou}, Christos and {Dvornik}, Andrej},
        title = "{Cosmological constraints from galaxy clustering and galaxy─galaxy lensing with extended SubHalo Abundance Matching}",
      journal = {\mnras},
         year = 2026,
        month = feb,
       volume = {545},
       number = {4},
          eid = {staf2143},
        pages = {staf2143},
          doi = {10.1093/mnras/staf2143},
archivePrefix = {arXiv},
       eprint = {2507.01601},
 primaryClass = {astro-ph.CO},
       adsurl = {https://ui.adsabs.harvard.edu/abs/2026MNRAS.545f2143M}
}

@ARTICLE{Angulo&White,
       author = {{Angulo}, R.~E. and {White}, S.~D.~M.},
        title = "{One simulation to fit them all - changing the background parameters of a cosmological N-body simulation}",
      journal = {\mnras},
         year = 2010,
        month = jun,
       volume = {405},
       number = {1},
        pages = {143-154},
          doi = {10.1111/j.1365-2966.2010.16459.x},
archivePrefix = {arXiv},
       eprint = {0912.4277},
 primaryClass = {astro-ph.CO},
       adsurl = {https://ui.adsabs.harvard.edu/abs/2010MNRAS.405..143A}
}

@ARTICLE{zhai_etal19_aemulus3,
       author = {{Zhai}, Zhongxu and {Tinker}, Jeremy L. and {Becker}, Matthew R. and {DeRose}, Joseph and {Mao}, Yao-Yuan and {McClintock}, Thomas and {McLaughlin}, Sean and {Rozo}, Eduardo and {Wechsler}, Risa H.},
        title = "{The Aemulus Project. III. Emulation of the Galaxy Correlation Function}",
      journal = {\apj},
         year = 2019,
        month = mar,
       volume = {874},
       number = {1},
          eid = {95},
        pages = {95},
          doi = {10.3847/1538-4357/ab0d7b},
archivePrefix = {arXiv},
       eprint = {1804.05867},
 primaryClass = {astro-ph.CO},
       adsurl = {https://ui.adsabs.harvard.edu/abs/2019ApJ...874...95Z}
}

@ARTICLE{2019ApJ...875...69D,
       author = {{DeRose}, Joseph and {Wechsler}, Risa H. and {Tinker}, Jeremy L. and {Becker}, Matthew R. and {Mao}, Yao-Yuan and {McClintock}, Thomas and {McLaughlin}, Sean and {Rozo}, Eduardo and {Zhai}, Zhongxu},
        title = "{The AEMULUS Project. I. Numerical Simulations for Precision Cosmology}",
      journal = {\apj},
         year = 2019,
        month = apr,
       volume = {875},
       number = {1},
          eid = {69},
        pages = {69},
          doi = {10.3847/1538-4357/ab1085},
archivePrefix = {arXiv},
       eprint = {1804.05865},
 primaryClass = {astro-ph.CO},
       adsurl = {https://ui.adsabs.harvard.edu/abs/2019ApJ...875...69D}
}

@ARTICLE{CEMBOSS1,
       author = {{Lange}, Johannes U. and {Hearin}, Andrew P. and {Leauthaud}, Alexie and {van den Bosch}, Frank C. and {Guo}, Hong and {DeRose}, Joseph},
        title = "{Five per cent measurements of the growth rate from simulation-based modelling of redshift-space clustering in BOSS LOWZ}",
      journal = {\mnras},
         year = 2022,
        month = jan,
       volume = {509},
       number = {2},
        pages = {1779-1804},
          doi = {10.1093/mnras/stab3111},
archivePrefix = {arXiv},
       eprint = {2101.12261},
 primaryClass = {astro-ph.CO},
       adsurl = {https://ui.adsabs.harvard.edu/abs/2022MNRAS.509.1779L}
}

@ARTICLE{CEMBOSS2,
       author = {{Lange}, Johannes U. and {Hearin}, Andrew P. and {Leauthaud}, Alexie and {van den Bosch}, Frank C. and {Xhakaj}, Enia and {Guo}, Hong and {Wechsler}, Risa H. and {DeRose}, Joseph},
        title = "{Constraints on S$_{8}$ from a full-scale and full-shape analysis of redshift-space clustering and galaxy-galaxy lensing in BOSS}",
      journal = {\mnras},
         year = 2023,
        month = apr,
       volume = {520},
       number = {4},
        pages = {5373-5393},
          doi = {10.1093/mnras/stad473},
archivePrefix = {arXiv},
       eprint = {2301.08692},
 primaryClass = {astro-ph.CO},
       adsurl = {https://ui.adsabs.harvard.edu/abs/2023MNRAS.520.5373L}
}

@ARTICLE{SDSSCiC,
       author = {{Wang}, Kuan and {Mao}, Yao-Yuan and {Zentner}, Andrew R. and {Guo}, Hong and {Lange}, Johannes U. and {van den Bosch}, Frank C. and {Mezini}, Lorena},
        title = "{Evidence of galaxy assembly bias in SDSS DR7 galaxy samples from count statistics}",
      journal = {\mnras},
         year = 2022,
        month = nov,
       volume = {516},
       number = {3},
        pages = {4003-4024},
          doi = {10.1093/mnras/stac2465},
archivePrefix = {arXiv},
       eprint = {2204.05332},
 primaryClass = {astro-ph.GA},
       adsurl = {https://ui.adsabs.harvard.edu/abs/2022MNRAS.516.4003W}
}

@ARTICLE{DESICiC,
       author = {{Pearl}, Alan N. and {Zentner}, Andrew R. and {Newman}, Jeffrey A. and {Bezanson}, Rachel and {Wang}, Kuan and {Moustakas}, John and {Aguilar}, Jessica N. and {Ahlen}, Steven and {Brooks}, David and {Claybaugh}, Todd and {Cole}, Shaun and {Dawson}, Kyle and {de la Macorra}, Axel and {Doel}, Peter and {Forero-Romero}, Jamie E. and {Gontcho A Gontcho}, Satya and {Honscheid}, Klaus and {Landriau}, Martin and {Manera}, Marc and {Martini}, Paul and {Meisner}, Aaron and {Miquel}, Ramon and {Nie}, Jundan and {Percival}, Will and {Prada}, Francisco and {Rezaie}, Mehdi and {Rossi}, Graziano and {Sanchez}, Eusebio and {Schubnell}, Michael and {Tarl{\'e}}, Gregory and {Weaver}, Benjamin A. and {Zhou}, Zhimin},
        title = "{The DESI One-percent Survey: Evidence for Assembly Bias from Low-redshift Counts-in-cylinders Measurements}",
      journal = {\apj},
         year = 2024,
        month = mar,
       volume = {963},
       number = {2},
          eid = {116},
        pages = {116},
          doi = {10.3847/1538-4357/ad1ffd},
archivePrefix = {arXiv},
       eprint = {2309.08675},
 primaryClass = {astro-ph.CO},
       adsurl = {https://ui.adsabs.harvard.edu/abs/2024ApJ...963..116P}
}

@ARTICLE{vdBosch2005,
       author = {{van den Bosch}, Frank C. and {Tormen}, Giuseppe and {Giocoli}, Carlo},
        title = "{The mass function and average mass-loss rate of dark matter subhaloes}",
      journal = {\mnras},
         year = 2005,
        month = may,
       volume = {359},
       number = {3},
        pages = {1029-1040},
          doi = {10.1111/j.1365-2966.2005.08964.x},
archivePrefix = {arXiv},
       eprint = {astro-ph/0409201},
 primaryClass = {astro-ph},
       adsurl = {https://ui.adsabs.harvard.edu/abs/2005MNRAS.359.1029V}
}

@ARTICLE{Gao2007,
       author = {{Gao}, Liang and {White}, Simon D.~M.},
        title = "{Assembly bias in the clustering of dark matter haloes}",
      journal = {\mnras},
         year = 2007,
        month = apr,
       volume = {377},
       number = {1},
        pages = {L5-L9},
          doi = {10.1111/j.1745-3933.2007.00292.x},
archivePrefix = {arXiv},
       eprint = {astro-ph/0611921},
 primaryClass = {astro-ph},
       adsurl = {https://ui.adsabs.harvard.edu/abs/2007MNRAS.377L...5G}
}

@ARTICLE{Diemer&Kravtsov2014,
       author = {{Diemer}, Benedikt and {Kravtsov}, Andrey V.},
        title = "{Dependence of the Outer Density Profiles of Halos on Their Mass Accretion Rate}",
      journal = {\apj},
         year = 2014,
        month = jul,
       volume = {789},
       number = {1},
          eid = {1},
        pages = {1},
          doi = {10.1088/0004-637X/789/1/1},
archivePrefix = {arXiv},
       eprint = {1401.1216},
 primaryClass = {astro-ph.CO},
       adsurl = {https://ui.adsabs.harvard.edu/abs/2014ApJ...789....1D}
}

@ARTICLE{zentner_etal14_assembias,
       author = {{Zentner}, Andrew R. and {Hearin}, Andrew P. and {van den Bosch}, Frank C.},
        title = "{Galaxy assembly bias: a significant source of systematic error in the galaxy-halo relationship}",
      journal = {\mnras},
         year = 2014,
        month = oct,
       volume = {443},
       number = {4},
        pages = {3044-3067},
          doi = {10.1093/mnras/stu1383},
archivePrefix = {arXiv},
       eprint = {1311.1818},
 primaryClass = {astro-ph.CO},
       adsurl = {https://ui.adsabs.harvard.edu/abs/2014MNRAS.443.3044Z}
}

@ARTICLE{Reddick2013,
       author = {{Reddick}, Rachel M. and {Wechsler}, Risa H. and {Tinker}, Jeremy L. and {Behroozi}, Peter S.},
        title = "{The Connection between Galaxies and Dark Matter Structures in the Local Universe}",
      journal = {\apj},
         year = 2013,
        month = jul,
       volume = {771},
       number = {1},
          eid = {30},
        pages = {30},
          doi = {10.1088/0004-637X/771/1/30},
archivePrefix = {arXiv},
       eprint = {1207.2160},
 primaryClass = {astro-ph.CO},
       adsurl = {https://ui.adsabs.harvard.edu/abs/2013ApJ...771...30R}
}

@ARTICLE{Watson2013,
       author = {{Watson}, Douglas F. and {Conroy}, Charlie},
        title = "{The Strikingly Similar Relation between Satellite and Central Galaxies and their Dark Matter Halos since z = 2}",
      journal = {\apj},
         year = 2013,
        month = aug,
       volume = {772},
       number = {2},
          eid = {139},
        pages = {139},
          doi = {10.1088/0004-637X/772/2/139},
archivePrefix = {arXiv},
       eprint = {1301.4497},
 primaryClass = {astro-ph.CO},
       adsurl = {https://ui.adsabs.harvard.edu/abs/2013ApJ...772..139W}
}

@ARTICLE{MDPL2,
       author = {{Klypin}, Anatoly and {Yepes}, Gustavo and {Gottl{\"o}ber}, Stefan and {Prada}, Francisco and {He{\ss}}, Steffen},
        title = "{MultiDark simulations: the story of dark matter halo concentrations and density profiles}",
      journal = {\mnras},
         year = 2016,
        month = apr,
       volume = {457},
       number = {4},
        pages = {4340-4359},
          doi = {10.1093/mnras/stw248},
archivePrefix = {arXiv},
       eprint = {1411.4001},
 primaryClass = {astro-ph.CO},
       adsurl = {https://ui.adsabs.harvard.edu/abs/2016MNRAS.457.4340K}
}

@ARTICLE{BerlindWeinberg2002,
       author = {{Berlind}, Andreas A. and {Weinberg}, David H.},
        title = "{The Halo Occupation Distribution: Toward an Empirical Determination of the Relation between Galaxies and Mass}",
      journal = {\apj},
         year = 2002,
        month = aug,
       volume = {575},
       number = {2},
        pages = {587-616},
          doi = {10.1086/341469},
archivePrefix = {arXiv},
       eprint = {astro-ph/0109001},
 primaryClass = {astro-ph},
       adsurl = {https://ui.adsabs.harvard.edu/abs/2002ApJ...575..587B}
}

\appendix 

\section{Choice of interpolation for SHAM} \label{app1}

In this work, we employ a SHAM model that includes one additional parameter beyond the standard SHAM approach to introduce a dependence beyond $M_{\rm peak}$, allowing a smooth interpolation between two halo proxies.
This is inspired by the similar approach of \cite{Lehmann2017} (hereafter L17), which we have modified slightly to expand the range of clustering amplitudes the model can predict. 
In this Appendix, we investigate the impact of different choices of halo proxies, particularly those related to the proxies used by L17.

The L17 model proposes a generalized halo proxy
\begin{equation}
    v_{\alpha} = v_{\rm vir}\Big(\frac{v_{\rm max}}{v_{\rm vir}}\Big)^{\alpha}, 
\end{equation}
where $v_{\rm max}$ is the maximal circular velocity and $v_{\rm vir}$ is the virial velocity of the halo. The authors note that this allows an adjustable dependence on halo concentration through the additional parameter $\alpha$, as the ratio $v_{\rm max}/v_{\rm vir}$ can be understood as a proxy for concentration. They evaluate this proxy at the epoch of $M_{\rm peak}$, $a_{M_{\rm peak}}$, for each halo, and propose that this parametrization allows the model to be an interpolation between the two most commonly used SHAM proxies: the peak halo mass $M_{\rm peak}$ and the peak halo maximal circular velocity $v_{\rm peak}$. Specifically, when $\alpha=0$, ranking with $v_{\alpha=0}$ is equivalent to ranking with $M_{\rm {peak}}$, and when $\alpha=1$, ranking with $v_{\alpha=1}$ is equivalent to ranking with $v_{\rm max}$ at $M_{\rm {peak}}$, which yields nearly identical galaxy samples as ranking with $v_{\rm {peak}}$. 

A subtlety arises in the evaluation of $v_{\rm{vir}}$ at $a_{M_\text{peak}}$, as the definition of $v_{\rm{vir}}$ depends on the virial overdensity $\Delta_{\rm{vir}}$ and the critical density $\rho_{\rm crit}$ in addition to halo mass, 
\begin{equation}
    v_{\text{vir}} = \Big(\frac{4\pi}{3}\Delta_{\text{vir}}\rho_{\text{crit}}G^{3} \Big)^{1/6}M_{\text{vir}}^{1/3},
\end{equation}
where $G$ is the gravitational constant, and the density threshold $\rho_{\rm vir}(a) \equiv \Delta_{\text{vir}}(a)\rho_{\text{crit}}(a)$ evolves strongly with the scale factor: it is larger in the early universe so that an object of fixed mass has higher virial velocity at higher redshift. Thus, $v_{\text{vir}}$ at $a_{M_{\rm peak}}$ could be the actual virial velocity of the halo at the epoch of $M_{\rm peak}$ adopting the density threshold at the same epoch, 
 \begin{equation} \label{Eq: actual}
    v_{\text{vir}} (a_{M_{\text{peak}}}) = \Big(\frac{4\pi}{3}\rho_{\text{vir}}(a = a_{M_{\rm peak}}) G^{3} \Big)^{1/6}M_{\text{peak}}^{1/3};
\end{equation}
or we can remove the redshift evolution of the density threshold by defining
\begin{equation} \label{Eq: L17}
    v'_{\text{vir}} (a_{M_{\text{peak}}}) = \Big(\frac{4\pi}{3} \rho_{\text{vir}}(a = 1) G^{3} \Big)^{1/6}M_{\text{peak}}^{1/3},
\end{equation}
so that this quantity serves only as a unit conversion of $M_{\rm peak}$, and thus ranking with $v_{\alpha=0}= v'_{\text{vir}} (a_{M_{\text{peak}}})$ is reduced to be exactly equivalent to ranking with $M_{\rm peak}$.
This definition $v_{\alpha=0} = v'_{\text{vir}}(a_{M_{\text{peak}}})$ is the one intended and used by L17 (private communication). 
For brevity, we write $v_{\text{vir}}$ and $v'_{\text{vir}} $ for $v_{\text{vir}} (a_{M_{\text{peak}}})$ and $v'_{\text{vir}} (a_{M_{\text{peak}}})$ in what follows.

\begin{figure*}
    \centering
    \includegraphics[width=0.85\columnwidth]{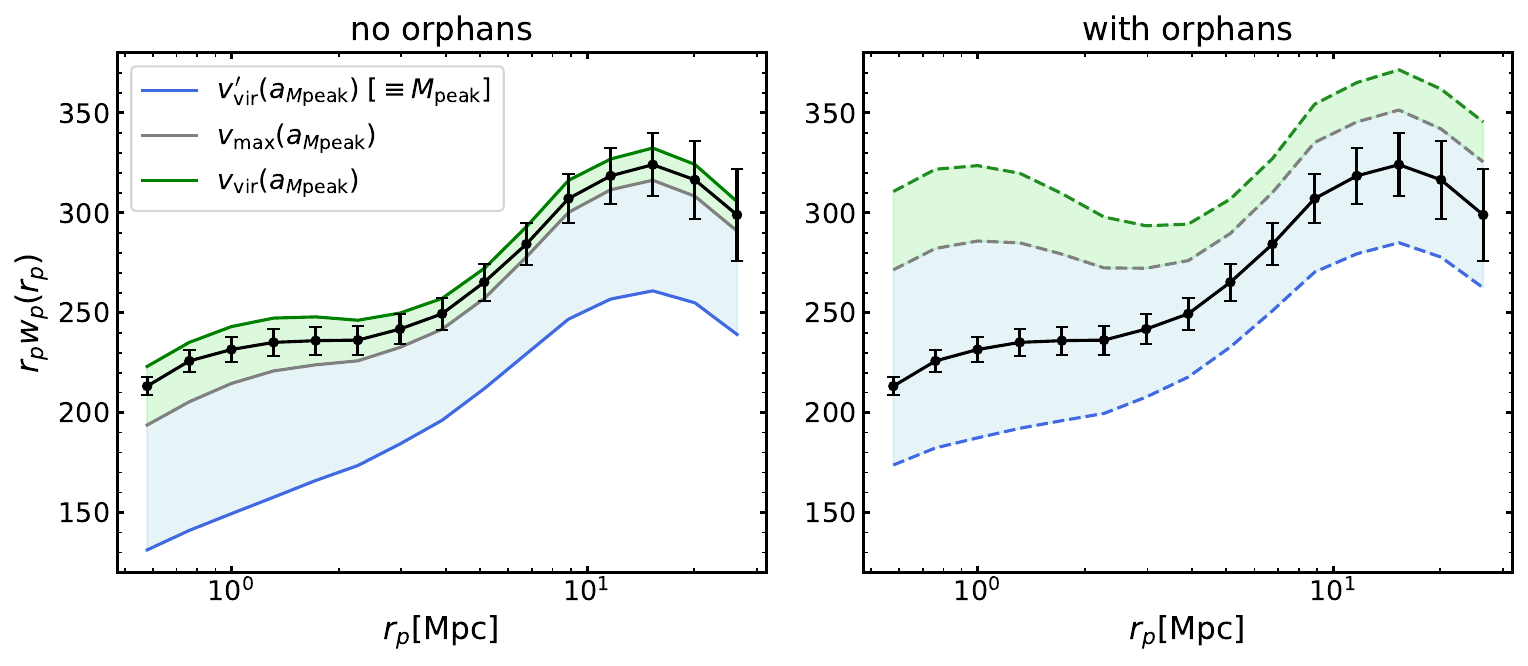}
    \caption{Projected correlation functions predicted by SHAM models using different halo proxies at fixed scatter. Left: the model built on the halo catalog without orphans; right: the same model built on halo catalog including orphans. The target \um\ measurement is shown in black.
    The blue shaded region marks the interpolation range between $M_{\rm peak}$ and $v_{\rm peak}$ (reported by L17), and the green shaded region shows the interpolation range between $v_{\text{vir}}(a_{M_{\text{peak}}})$ and $v_{\rm peak}$. The SHAM implementation adopted in this work spans the full interpolation range, combining both shaded regions.} 
    \label{fig: L17_test}
\end{figure*}

Because $a_{M_\text{peak}}$ varies across the halo population, the two definitions of virial velocity can produce different clustering amplitudes when used as the halo proxy.
As shown in Fig.~\ref{fig: L17_test}, abundance matching by $v'_{\rm vir}$ leads to lower clustering amplitude than abundance matching by $v_{\rm max}(a_{M_{\rm peak}})$, but abundance matching by $v_{\text{vir}}$ predicts higher clustering amplitudes.
Since $\alpha$ enables a smooth interpolation between two different halo proxies, this leads to different interpolation ranges allowed by the models, shown by the shaded regions in Fig.~\ref{fig: L17_test}.
In particular, when using the actual virial velocity at $a_{M_{\rm peak}}$ (Eq.~\ref{Eq: actual}), both the clustering amplitude and the satellite fraction decrease with larger $\alpha$, while these quantities increase with $\alpha$ when adopting the definition $v'_{\text{vir}}$ (Eq.~\ref{Eq: L17}). 

These differences are closely connected to the physical interpretation of the models, as the halo proxies differ in how they rank the halos at fixed $M_{\rm peak}$. 
Matching by $v'_{\rm vir}$ is equivalent to matching by $M_{\rm peak}$ and gives the lowest clustering. Matching by $v_{\rm vir}$ assigns satellites higher stellar masses than centrals at the same $M_{\rm peak}$, because satellites tend to reach $M_{\rm peak}$ earlier than centrals and the virial density threshold is larger at higher redshifts. This leads to an increase in satellite fraction across the sample, which increases the clustering amplitude. Matching by $v_{\rm max}(a_{M_{\rm peak}})$ increases the satellite fraction relative to matching by $v'_{\rm vir}$, but less than by $v_{\rm vir}$, giving intermediate clustering.
The opposite trends of the satellite fraction and the clustering amplitude with $\alpha$ then follow from which proxy is adopted at $\alpha=0$. 
L17 interpret $\alpha$ as tuning the dependence on halo concentration in their model, with higher-concentration halos ranked higher at larger $\alpha$, boosting the satellite fraction and increasing the clustering amplitude. We note, however, that the ratio being tuned, $v_{\rm max}(a_{M_{\rm peak}})/v'_{\rm vir}$, differs from the conventional concentration proxy, $v_{\rm max}/v_{\rm vir}$ with both evaluated at the same epoch, in that it combines each halo's concentration at $M_{\rm peak}$ and the amount of evolution in the virial density threshold between $a_{M_{\rm peak}}$ and the present day. We find that this ratio correlates much more weakly with the present-day concentration than $v_{\rm max}/v_{\rm vir}$ evaluated at the present day.

In this work, we choose an alternative halo proxy parametrization $v_\alpha = v'_{\rm vir} (v_{\rm vir}/v'_{\rm vir})^\alpha$ (see Section~\ref{sec: SHAM}).
This parametrization combines the interpolation ranges allowed by both definitions of $v_{\rm vir}$, offering greater flexibility in the SHAM model compared to both L17 and the standard approach using individual halo proxies.
Here the ratio adjusted by $\alpha$, $v_{\rm vir}/v'_{\rm vir}$, directly captures the amount of evolution in the virial density threshold $\rho_{\text{vir}}$ between each halo's $a_{M_{\rm peak}}$ and the present day, and depends on $a_{M_{\rm peak}}$ alone monotonically.
The model thus introduces a clean adjustable dependence on the epoch of peak mass, which correlates with halo formation time, with larger $\alpha$ favoring halos that reach $M_{\rm peak}$ earlier.
Since subhalos tend to reach their peak mass earlier than host halos, increasing $\alpha$ increases their proxy values relative to centrals at fixed $M_{\rm peak}$. 
This parameterization therefore allows satellites to follow a different stellar-mass-to-halo-mass relation than centrals: the two follow the same relation for $\alpha=0$ and satellites receive larger stellar masses as $\alpha$ increases. 
As a result, the model can effectively tune the satellite fraction, and hence the clustering amplitude. 

Fig.~\ref{fig: L17_test} compares the interpolation ranges allowed by this SHAM model when built on halo catalogs with and without orphans. 
When built on the halo catalog without orphans, we observe that the effective tuning of satellite fraction of this SHAM parameterization allows it to emulate the effect of orphan modeling in \um. 
When built on the halo catalog including orphans, the interpolation range is noticeably wider. This makes sense as the orphan population is preferentially older and associates with host halos that are more strongly clustered, allowing extra flexibility in the clustering properties. 
This wider interpolation range leads to the weaker cosmological constraint from the with-orphans model in Section~\ref{subsec: cos_sham}.

\section{Recovery Test for SHAM} \label{app2}
\cite{CEM} demonstrates that the CEM method is able to recover cosmological information when assuming the HOD framework for the galaxy--halo connection. 
In this appendix, we investigate the ability of the SHAM framework to be applied as the galaxy--halo connection model to infer cosmology using CEM, in order to demonstrate the robustness of our analysis in this paper. 

\begin{figure*}
    \centering
    \includegraphics[width=0.85\columnwidth]{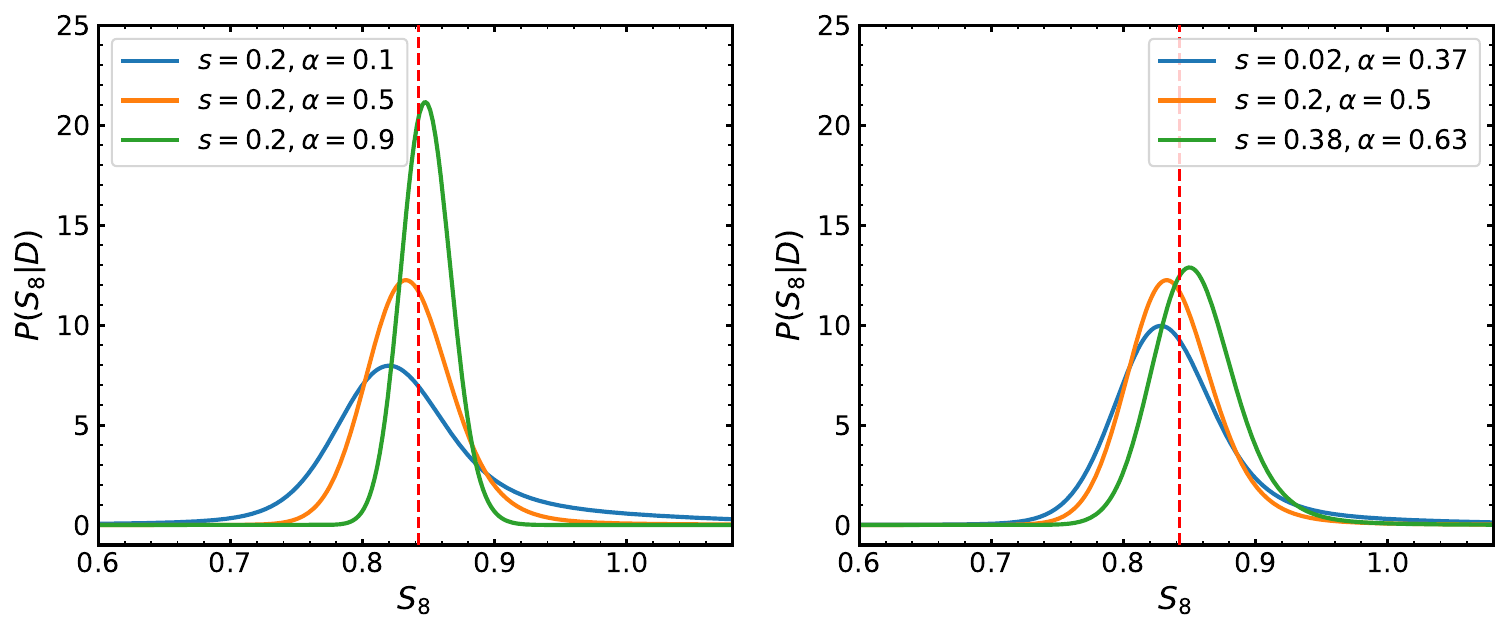}
    \caption{Cosmological constraints on $S_8$ from CEM with SHAM, when fiducial $w_p$ predicted by the same SHAM model using different parameter values are treated as the target data vector. The left panel demonstrates the recovery when the target SHAM model assumes a fixed scatter of $s=0.2$ and different values for $\alpha$; the right panel demonstrates the case when $s$ and $\alpha$ are tuned simultaneously for similar target clustering amplitudes. Cosmological constraints on $S_8$ are unbiased for all of the target SHAM-based data vectors investigated here.} 
    \label{fig: SHAM_recon}
\end{figure*}

We generate fiducial projected correlation functions with SHAM assuming different values of the two model parameters $s$ and $\alpha$ as the target data vectors, and repeat the CEM analysis to test whether the methodology is able to recover unbiased cosmology when the model used in the inference is the same as the one used to construct the target.
The left panel of Fig.~\ref{fig: SHAM_recon} shows the modeled posterior of $S_8$ when the scatter is fixed at $s=0.2$ and $\alpha$ is varied for different clustering amplitudes, while the right panel shows the modeled posterior when $s$ and $\alpha$ are tuned simultaneously to generate similar target clustering. For all target data vectors using different combinations of SHAM parameters, we see that CEM together with the SHAM framework is able to recover the input $S_8$, inferring unbiased cosmology. 

We note that the constraint on $S_8$ tightens as $\alpha$ increases at fixed $s$ from the left panel of Fig.~\ref{fig: SHAM_recon}. This is because the clustering amplitude increases for larger $\alpha$, while higher clustering is intrinsically harder to fit since more physical properties of the galaxy--halo connection, such as assembly bias, become important and present in the statistics, leaving narrower parameter space consistent with the measurement.
We further test this reasoning by tuning $s$ and $\alpha$ simultaneously to generate similar $w_p$ as the target data vectors. As shown in the right panel of Fig.~\ref{fig: SHAM_recon}, the cosmological constraints are approximately equivalent in these cases, indicating that the constraining power of CEM on cosmology is dependent on the complexity of galaxy formation encoded in the data vector. 

In summary, for a wide range of combinations of SHAM parameters, the CEM method assuming the SHAM framework is able to recover unbiased cosmological information, suggesting that the method is robust enough to apply to cosmological analysis targeting more complex galaxy samples in this work. 


\section{Galaxy--Halo Connection from SHAM with different orphan modeling} \label{app3}
As presented in the main text, both SHAM models built on halo catalogs with and without orphans are able to fit the target $w_p$ and offer unbiased cosmological constraints on $S_8$. However, they are intrinsically different models, as the underlying subhalo populations are certainly distinct. In this appendix, we demonstrate a difference between these two satellite models that is not reflected in the two-point summary statistic.

\begin{figure*}
    \centering
    \includegraphics[width=0.85\columnwidth]{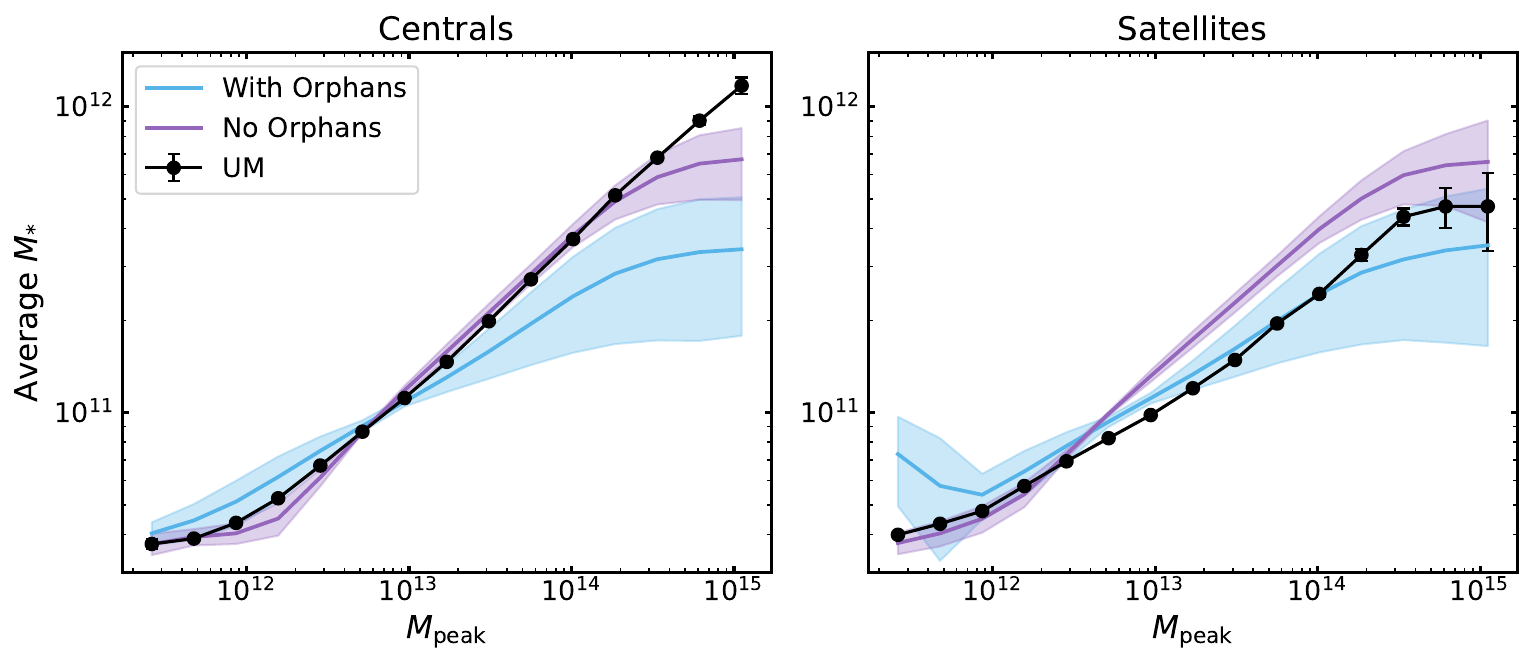}
    \caption{The stellar-mass to halo-mass relation for centrals (left) and satellites (right) predicted by the two SHAM models built on halo catalogs with and without orphans, compared to the measurements from \um. The solid curves show the mean of the posterior predictive distribution for each model, and the shaded regions are the $1\sigma$ interval; error bars on the \um\ measurements correspond to the standard error of the mean in each halo mass bin.
    The with-orphans and no-orphans SHAM models predict distinct stellar-mass to halo-mass relations, suggesting the inherent difference between the models.}
    \label{fig: SMHM}
\end{figure*}


Fig.~\ref{fig: SMHM} shows the stellar-mass to halo-mass relation (SHMR) for centrals and satellites predicted by the two SHAM models, compared to the actual measurements from \um. The solid curves show the mean of the posterior predictive distribution for the SHMR from each model, with shaded regions giving the $1\sigma$ ranges. Error bars for the \um\ measurements are estimated analytically as $\sigma/\sqrt{N}$, where $\sigma$ is the scatter in stellar mass and $N$ is the number of objects in each halo mass bin.
The two SHAM models predict noticeably different SHMRs, demonstrating that the with-orphans and no-orphans models are distinct even though they offer comparable predictions for two-point clustering.
For instance, the no-orphans model in general predicts higher satellite stellar masses than the with-orphans model, which is consistent with previous literature that satellite fractions can be boosted to the required level either by adding orphans or increasing the stellar mass at fixed subhalo mass \citep{Reddick2013, Watson2013, UniverseMachine}. 
The difference in the predicted SHMRs suggests that $w_p$ only weakly distinguishes between the orphan treatments for our galaxy sample, though this conclusion may not extend to lower stellar-mass samples or other simulation setups.
Meanwhile, more complex statistics are necessary to break this degeneracy. We attempt to evaluate several higher-order statistics in Section~\ref{sec: acm}, but reaching a quantitative conclusion would require more careful calibrations of the covariance matrix. 

Even though the summary statistic $w_p$ only weakly constrains the orphan population and the best-fit SHAM models do not infer exactly the correct galaxy--halo connection (e.g. as reflected in the predicted SHMRs), SHAM as the modeling framework successfully marginalizes over uncertainties in the galaxy--halo connection and supports unbiased cosmological inference.
We leave a comprehensive exploration of more complex summary statistics and validation tests to quantify degeneracies between cosmological parameters and galaxy biases for the future. 

\end{document}